\documentclass[aps,pra,showpacs,superscriptaddress,preprint]{revtex4}
\usepackage{amsmath}
\usepackage{graphicx}

\catcode`ð=\active
 \defð{\u{g}}
 \catcode`Ð=\active
 \defÐ{\u{G}}
 \catcode`Ý=\active
\defÝ{\. I}
 \catcode`ö=\active
\defö{\"{o}}
 \catcode`Ö=\active
 \defÖ{\"O}
 \catcode`ü=\active
 \defü{\"{u}}
 \catcode`Ü=\active
 \defÜ{\"{U}}
 \catcode`Þ=\active
\defÞ{\c{S}}
 \catcode`þ=\active
 \defþ{\c{s}}
 \catcode`ý=\active
 \defý{{\i}}
 \catcode`ç=\active
\defç{\d{c}}
 \catcode`Ç=\active
\defÇ{\d{C}}

\input{tcilatex}

\begin{document}

\title{An approximate $\kappa $ state solutions of the Dirac equation for
the generalized Morse potential under spin and pseudospin symmetry }
\author{Sameer M. Ikhdair}
\email[E-mail: ]{sikhdair@neu.edu.tr}
\affiliation{Physics Department, Near East University, Nicosia, N. Cyprus, Turkey}
\date{%
\today%
}

\begin{abstract}
By using an improved approximation scheme to deal with the centrifugal
(pseudo-centrifugal) term, we solve the Dirac equation for the generalized
Morse potential with arbitrary spin-orbit quantum number $\kappa .$ In the
presence of spin and pseudospin symmetry, the analytic bound state energy
eigenvalues and the associated upper- and lower-spinor components of two
Dirac particles are found by using the basic concepts of the
Nikiforov-Uvarov method. We study the special cases when $\kappa =\pm 1$ ($l=%
\widetilde{l}=0,$ $s$-wave), the non-relativistic limit and the limit when $%
\alpha $ becomes zero (Kratzer potential model). The present solutions are
compared with those obtained by other methods.

\end{abstract}

\maketitle

\section{Introduction}

The simplest modified Morse potential model suggested by Deng and Fan [1]
and related to the Manning-Rosen potential [2] (also called Eckart potential
by some authors [3]) is anharmonic potential defined by 
\begin{equation}
V(r)=D\left( 1-\frac{b}{e^{\alpha r}-1}\right) ^{2}\text{ , }b=e^{\alpha
r_{e}}-1,
\end{equation}%
where $r\in (0,\infty ),$ and the three positive parameters $D,$ $r_{e}$ and 
$\alpha $ denote the dissociation energy, the equilibrium inter-nuclear
distance and the range of the potential well, respectively. The above
potential is used to describe diatomic molecular energy spectra and
electromagnetic transitions and is the true internuclear potential in
diatomic molecules with the same behaviour for $r\rightarrow 0$ [4]$.$ The
above potential was called a generalized Morse potential (GMP) model and
illustrated in Figure 1 for various values of potential parameters. As
stated in Ref. [4], the Morse potential and the GMP are very close to each
other for large values of $r_{e}$ $(\alpha =1)$ in the regions $r\sim r_{e}$
and $r>r_{e}$ but are very different at $r\sim 0.$ Further, if the two
potentials are deep $(D\gg 1),$ they could be well approximated by a
harmonic oscillator in the region $r\sim r_{e}$ (see Fig.1 in [4]). To
describe the vibrational spectra of diatomic molecules the potential curve $%
V(r)$ is approximated by a sum of three Morse functions [5]. The approach
has the advantage of being more flexible than the simple Morse potential
while preserving the correct asymptotical $V(\infty )=const$ [5]$.$ The
potential model (1) is a special case of the five-parameter exponential-type
potential model [6,7]. The exact solvability of the $s$-wave ($l=0$ case)
bound state energy eigenvalues and eigenfunctions of the GMP is due to the
fact that it belongs to the class of the Eckart potential, a member of the
hypergeometric Natanzon potentials which can be solved algebraically by
means of $SO(2,2)$ symmetry algebra [4] and $SO(2,1)$ algebra [8]. The X-H
stretching motion in small molecules has been treated by the potential model
(1) [9]. Moreover, the approximated $l$-wave bound state solutions [10] of
the Schr\"{o}dinger equation has been solved using the conventional
approximation scheme suggested by Greene and Aldrich [11] to deal with the
centrifugal barrier term $l(l+1)/r^{2},$ singularity at $r=0.$ The exact
analytic expressions for matrix elements of positive integral powers of the
coordinate [12] and the quasi-one-dimensional system of DNA [13] have also
been studied.

To study the relativistic effects and corrections of the molecular Morse
potential [4], the Dirac equation has been solved for attractive scalar $%
S(r) $ and repulsive vector $V(r)$ Morse potentials under pseudospin
symmetry in the nuclear theory using the Pekeris approximation [14].
Recently, the approximate bound state solutions of the pseudospin and spin
symmetric Dirac equation with the GMP has been calculated using an improved
approximation scheme to deal with the centrifugal (pseudo-centrifugal) term
and by employing the basic concept of the supersymmetric shape invariance
formalism (cf. [15] and the references therein).

Many authors have investigated approximately the solution of the Dirac
equation with a few potential models such as the Hulth\'{e}n potential [16],
the Hulthen potential including Coulomb-like tensor potential [17], the
generalized Woods-Saxon potential [18,19], the\ Eckart \ potential [20], the
Morse potential [14,21], the P\"{o}schl-Teller potential [22], the
Manning-Rosen potential [2,23], the hyperbolic potential [24], the
Rosen-Morse potential [25], the pseudoharmonic potential [26], and the
Kratzer potential connected with an angle-dependent potential [27], etc
within the framework of various methods.

The diatomic molecular model consisting of nuclei having masses $m_{1}$ and $%
m_{2},$ the reduced mass is defined $\mu =m_{1}m_{2}/(m_{1}+m_{2})$ can be
included to the spin symmetry and the pseudospin symmetry concepts [28].
Ginocchio [29-31] showed that the spin symmetry occurs when the difference
potential between the vector potential $V(r)$ and scalar potential $S(r)$ is
a constant (i.e., $\Delta (r)=V(r)-S(r)=C_{s}=$ constant) and the pseudospin
symmetry occurs when the sum potential of the vector potential $V(r)$ and
scalar potential $S(r)$ is a constant (i.e., $\Sigma (r)=V(r)-S(r)=C_{ps}=$
constant). The spin symmetry concept [32] is particularly relevant for
mesons [33]. The pseudospin symmetry concept [34,35] in nuclear physics
refers to the quasi-degeneracy of single-nucleon doublets and can be
characterized with the non-relativistic quantum numbers $(n,l,j=l+1/2)$ and $%
(n-1,l+2,j=l+3/2),$ where $n,$ $l$ and $j$ are the single-nucleon radial,
orbital and total angular momentum quantum numbers, respectively. Alhaidari 
\textit{et al}. [36] investigated in detail the physical interpretation on
the three-dimensional Dirac equation in the cases of spin symmetry
limitation $\Delta (r)=0$ and pseudospin symmetry\ limitation $\Sigma (r)=0.$
In real nuclei, $\Sigma (r)\neq $ constant and pseudospin symmetry is only
an approximation. The quality of the pseudospin symmetry approximation
depends on the competition between the pseudo-centrifugal potential and the
pseudospin orbital potential [37].

Recently, in the framework of the spin symmetry $S(r)\sim V(r)$ and
pseudospin symmetry $S(r)\sim -V(r)$, the bound state energy eigenvalues and
associated upper- and lower-spinor wave functions are investigated by means
of the Nikiforov-Uvarov (NU) method [38]. We have approximately solved the
Dirac equation for the Rosen-Morse potential [25] with spin and pseudospin
symmetry for any $\kappa $ state and found the eigenvalue equation and
corresponding two-component spinors within the framework of an approximation
to the term proportional to $1/r^{2}.$ We have also solved the (3+1)
dimensional Dirac equation for a particle trapped in the spherically
symmetric generalized WS potential under the conditions of exact spin and
pseudospin symmetry combined with approximation for the spin-orbit
centrifugal (pseudo-centrifugal) term, and calculated the two-component
spinor wave functions and the energy eigenvalues for any arbitrary
spin-orbit $\kappa $ bound states [18].

In principle, the solution of the Schr\"{o}dinger equation for the GMP model
can be used to describe the motion of the nucleons in the mean field
produced by the interactions between nuclei. However, the Dirac equation
successfully merges quantum mechanics with special relativity and is
considered to be the natural transition to quantum field theory. It provides
a natural description of the electron spin, predicts the existence of
antimatter and is able to reproduce accurately the spectrum of the hydrogen
atom. It also predicts some peculiar phenomena such as Klein's paradox and
unexpected quivering motion of free relativistic quantum particle which are
key examples for understanding relativistic quantum effects, but are
difficult to observe in real particles. In order to to understand the origin
of spin and pseudospin symmetry, we need to take into consideration the
motion of the nucleons in a relativistic mean field and consider the Dirac
equation [31]. We attempt to study the approximate solution of the Dirac
equation for the GMP with non-zero spin-orbit quantum numbers $\kappa $ by
employing an improved approximation scheme (see e.g., [17,39,40] and the
references therein) to deal with the spin-orbit centrifugal, $\kappa (\kappa
+1)/r^{2}$ (pseudo-centrifugal, $\kappa (\kappa -1)/r^{2}$) barrier term.
The inter-relation between the GMP with the Oscillator and Morse potentials
investigated in Ref. [4] and the recent relativistic treatments of the Morse
and Oscillator potentials in [14,26] are some motivations for the present
study. The simple transformation of the Dirac equation into the Schr\"{o}%
dinger-like equation and the success in studying the approximated bound
state solutions of the Dirac equation with various potential models
[14,17,18,25,26] also make the solution possible.

We tend to show that the new scheme of parametric generalization of the NU
method [41] given in Appendix A is a powerful tool for solving a second
order differential equation by turning it into a hypergeometric type
equation [38]. The advantage of employing the NU method, in the present
work, is that it can be used to find the bound state energy spectra and the
corresponding spinor wave functions under the condition of spin symmetry and
pseudospin symmetry concept for any $\kappa $ state in a very simple way.

This paper is organized as follows. In section 2, we investigate the bound
state energy equation and the corresponding two-component spinor wave
functions under the condition of spin symmetry and pseudospin symmetry
concept for the GMP model by employing a parametric generalization of the NU
method. In section 3, we study some special cases like the $s(\widetilde{s})$%
-wave cases $\kappa =\pm 1$ ($l=\widetilde{l}=0$), the non-relativistic
limit and the $\alpha \rightarrow 0$ (Kratzer potential). In section 4, we
present some numerical results to the non-relativistic and relativistic
numerical energy levels for GMP and Kratzer models. The relevant conclusions
are given in section 5.

\section{Dirac bound state solutions}

The Dirac equation for fermionic massive spin-$1/2$ particles moving in an
attractive scalar potential $S(r)$ and a repulsive vector potential $V(r)$
is given by [42] 
\begin{equation}
\left[ c\mathbf{\alpha }\cdot \mathbf{p+\beta }\left( Mc^{2}+S(r)\right)
+V(r)-E\right] \psi _{n\kappa }(\mathbf{r})=0,\text{ }\psi _{n\kappa }(%
\mathbf{r})=\psi (r,\theta ,\phi ),
\end{equation}%
where $E$ is the relativistic energy of the system, $m$ is the mass of a
particle, $\mathbf{p}=-i\hbar \mathbf{\nabla }$ is the momentum operator,
and $\mathbf{\alpha }$ and $\mathbf{\beta }$ are $4\times 4$ Dirac matrices,
i.e.,%
\begin{equation}
\mathbf{\alpha =}\left( 
\begin{array}{cc}
0 & \mathbf{\sigma }_{i} \\ 
\mathbf{\sigma }_{i} & 0%
\end{array}%
\right) ,\text{ }\mathbf{\beta =}\left( 
\begin{array}{cc}
\mathbf{I} & 0 \\ 
0 & -\mathbf{I}%
\end{array}%
\right) ,\text{ }\sigma _{1}\mathbf{=}\left( 
\begin{array}{cc}
0 & 1 \\ 
1 & 0%
\end{array}%
\right) ,\text{ }\sigma _{2}\mathbf{=}\left( 
\begin{array}{cc}
0 & -i \\ 
i & 0%
\end{array}%
\right) ,\text{ }\sigma _{3}\mathbf{=}\left( 
\begin{array}{cc}
1 & 0 \\ 
0 & -1%
\end{array}%
\right) ,
\end{equation}%
where $\mathbf{I}$ denotes the $2\times 2$ identity matrix and $\mathbf{%
\sigma }_{i}$ are the three-vector Pauli spin matrices. For a spherical
symmetrical nuclei, the total angular momentum operator of the nuclei $%
\mathbf{J}$ and spin-orbit matrix operator $\mathbf{K}=-\mathbf{\beta }%
\left( \mathbf{\sigma }\cdot \mathbf{L}+\mathbf{I}\right) $ commute with the
Dirac Hamiltonian, where $\mathbf{L}$ is the orbital angular momentum
operator. The spinor wave functions can be classified according to the
radial quantum number $n$ and the spin-orbit quantum number $\kappa $ and
can be written using the Pauli-Dirac representation in the following forms:%
\begin{equation}
\psi _{n\kappa }(\mathbf{r})=\left( 
\begin{array}{c}
f_{n\kappa }(\mathbf{r}) \\ 
g_{n\kappa }(\mathbf{r})%
\end{array}%
\right) =\frac{1}{r}\left( 
\begin{array}{c}
F_{n\kappa }(r)Y_{jm\kappa }^{l}(\widehat{r}) \\ 
iG_{n\kappa }(r)Y_{jm(-\kappa )}^{\widetilde{l}}(\widehat{r})%
\end{array}%
\right) ,
\end{equation}%
where the upper- and lower-spinor components $F_{n\kappa }(r)$ and $%
G_{n\kappa }(r)$ are the real square-integral radial wave functions, $%
Y_{jm\kappa }^{l}(\widehat{r})$ and $Y_{jm(-\kappa )}^{\widetilde{l}}(%
\widehat{r})$ are the spin spherical harmonic functions coupled to the total
angular momentum $j$ and it's projection $m$ on the $z$ axis and $\kappa
\left( \kappa +1\right) =l(l+1)$ and $\kappa \left( \kappa -1\right) =%
\widetilde{l}(\widetilde{l}+1)$. The quantum number $\kappa $ is related to
the quantum numbers for spin symmetry $l$ and pseudospin symmetry $%
\widetilde{l}$ as%
\begin{equation}
\kappa =\left\{ 
\begin{array}{cccc}
-\left( l+1\right) =-\left( j+\frac{1}{2}\right) , & (s_{1/2},p_{3/2},\text{%
\textit{etc.}}), & \text{ }j=l+\frac{1}{2}, & \text{aligned spin }\left(
\kappa <0\right) , \\ 
+l=+\left( j+\frac{1}{2}\right) , & \text{ }(p_{1/2},d_{3/2},\text{\textit{%
etc.}}), & \text{ }j=l-\frac{1}{2}, & \text{unaligned spin }\left( \kappa
>0\right) ,%
\end{array}%
\right.
\end{equation}%
and the quasi-degenerate doublet structure can be expressed in terms of a
pseudospin angular momentum $\widetilde{s}=1/2$ and pseudo-orbital angular
momentum $\widetilde{l}$ which is defined as 
\begin{equation}
\kappa =\left\{ 
\begin{array}{cccc}
-\widetilde{l}=-\left( j+\frac{1}{2}\right) , & (s_{1/2},\text{ }p_{3/2},%
\text{ \textit{etc.}}), & j=\widetilde{l}-1/2, & \text{aligned spin }\left(
\kappa <0\right) , \\ 
+\left( \widetilde{l}+1\right) =+\left( j+\frac{1}{2}\right) , & \text{ }%
(d_{3/2},\text{ }f_{5/2},\text{ \textit{etc.}}), & \ j=\widetilde{l}+1/2, & 
\text{unaligned spin }\left( \kappa >0\right) ,%
\end{array}%
\right.
\end{equation}%
where $\kappa =\pm 1,\pm 2,\cdots .$ For example, ($1s_{1/2},0d_{3/2}$) and
(2p$_{3/2},1f_{5/2}$) can be considered as pseudospin doublets.

Upon direct substitution of Eq. (4) into Eq. (2), we can obtain two radial
coupled Dirac equations for the two spinor components as follows: 
\begin{subequations}
\begin{equation}
\left( \frac{d}{dr}+\frac{\kappa }{r}\right) F_{n\kappa }(r)=\left[
Mc^{2}+E_{n\kappa }-\Delta (r)\right] G_{n\kappa }(r),
\end{equation}%
\begin{equation}
\left( \frac{d}{dr}-\frac{\kappa }{r}\right) G_{n\kappa }(r)=\left[
Mc^{2}-E_{n\kappa }+\Sigma (r)\right] F_{n\kappa }(r),
\end{equation}%
where $\Delta (r)=V(r)-S(r)$ and $\Sigma (r)=V(r)+S(r)$ are the difference
and sum potentials, respectively.

Under the spin symmetry ( i.e., $\Delta (r)=C_{s}=$ constant), one can
eliminate $G_{n\kappa }(r)$ in Eq. (7a), with the aid of Eq. (7b), to obtain
a second-order differential equation for the upper-spinor component as
follows: 
\end{subequations}
\begin{equation*}
\left[ -\frac{d^{2}}{dr^{2}}+\frac{\kappa \left( \kappa +1\right) }{r^{2}}+%
\frac{1}{\hbar ^{2}c^{2}}\left( Mc^{2}+E_{n\kappa }-C_{s}\right) \Sigma (r)%
\right] F_{n\kappa }(r)
\end{equation*}%
\begin{equation}
=\frac{1}{\hbar ^{2}c^{2}}\left( E_{n\kappa }^{2}-M^{2}c^{4}+C_{s}\left(
Mc^{2}-E_{n\kappa }\right) \right) F_{n\kappa }(r),
\end{equation}%
and the lower-spinor component is obtained from Eq. (7a) as%
\begin{equation}
G_{n\kappa }(r)=\frac{1}{Mc^{2}+E_{n\kappa }-C_{s}}\left( \frac{d}{dr}+\frac{%
\kappa }{r}\right) F_{n\kappa }(r),
\end{equation}%
where $E_{n\kappa }\neq -Mc^{2},$ only real positive energy states exist
when $C_{s}=0$ (exact spin symmetry). On the other hand, under the
pseudospin symmetry ( i.e., $\Sigma (r)=C_{ps}=$ constant), one can
eliminate $F_{n\kappa }(r)$ in Eq. (7b), with the aid of Eq. (7a), to obtain
a second-order differential equation for the lower-spinor component as
follows:%
\begin{equation*}
\left[ -\frac{d^{2}}{dr^{2}}+\frac{\kappa \left( \kappa -1\right) }{r^{2}}-%
\frac{1}{\hbar ^{2}c^{2}}\left( Mc^{2}-E_{n\kappa }+C_{ps}\right) \Delta (r)%
\right] G_{n\kappa }(r)
\end{equation*}%
\begin{equation}
=\frac{1}{\hbar ^{2}c^{2}}\left( E_{n\kappa }^{2}-M^{2}c^{4}-C_{ps}\left(
Mc^{2}+E_{n\kappa }\right) \right) G_{n\kappa }(r),
\end{equation}%
and the upper-spinor component $F_{n\kappa }(r)$ is obtained from Eq. (7b) as%
\begin{equation}
F_{n\kappa }(r)=\frac{1}{Mc^{2}-E_{n\kappa }+C_{ps}}\left( \frac{d}{dr}-%
\frac{\kappa }{r}\right) G_{n\kappa }(r),
\end{equation}%
where $E_{n\kappa }\neq Mc^{2},$ only real negative energy states exist when 
$C_{ps}=0$ (exact pseudospin symmetry). It is worthy to note that the
reality and finiteness of our solutions demand that the upper and lower
radial components should satisfy the essential boundary conditions: $%
F_{n\kappa }(0)=G_{n\kappa }(0)=0$ and $F_{n\kappa }(\infty )=G_{n\kappa
}(\infty )=0.$

\subsection{Spin symmetry solutions of the GMP model}

At first, we investigate the spin symmetry by taking the $\Sigma
(r)=2V(r)\rightarrow V_{GMP}(r)$ as mentioned in Ref. [43] enables one to
reduce the resulting relativistic solutions into their non-relativistic
limit under appropriate transformations. From Eq. (8), we can see that the
energy eigenvalues, $E_{n\kappa },$ depend only on $n$ and $l,$ i.e., $%
E_{n\kappa }=E(n,l(l+1)).$ For $l\neq 0,$ the states with $j=l\pm 1/2$ are
degenerate. This is a $SU(2)$ spin symmetry. Following Refs. [22-25], we
impose the GMP [1,4] as the $\Sigma (r),$ i.e.,%
\begin{equation}
\Sigma (r)=D\left( 1-\frac{b}{e^{\alpha r}-1}\right) ^{2},
\end{equation}%
leads us to obtain a Schr\"{o}dinger-like equation in the form:%
\begin{equation}
\left[ \frac{d^{2}}{dr^{2}}-\frac{\kappa \left( \kappa +1\right) }{r^{2}}%
-\alpha ^{2}\nu _{1}^{2}\left( 1-\frac{be^{-\alpha r}}{1-e^{-\alpha r}}%
\right) ^{2}+\alpha ^{2}\omega _{1}^{2}\right] F_{n\kappa }(r)=0
\end{equation}%
\begin{equation}
\nu _{1}^{2}=\frac{1}{\alpha ^{2}\hbar ^{2}c^{2}}\left( Mc^{2}+E_{n\kappa
}-C_{s}\right) D,\text{ \ }\omega _{1}^{2}=\frac{1}{\alpha ^{2}\hbar
^{2}c^{2}}\left[ E_{n\kappa }^{2}-M^{2}c^{4}+C_{s}\left( Mc^{2}-E_{n\kappa
}\right) \right] ,
\end{equation}%
where $\kappa \left( \kappa +1\right) =l\left( l+1\right) ,$ $\kappa =l$ for 
$\kappa <0$ and $\kappa =-\left( l+1\right) $ for $\kappa >0.$ The exact
solution of the above equation is possible only for the $s$-wave case $%
(\kappa =-1)$ due to the centrifugal term $\kappa \left( \kappa +1\right)
/r^{2}.$ However, if $\kappa $ is not too large, the case of the vibrations
of small amplitude about the minimum, we attempt to use the following
improved new approximation scheme to deal with the centrifugal (pseudo
centrifugal) term, near the minimum point $r=r_{e}$ $(i.e.,$ $x=0),$ (cf.
Refs. [17,40,44-48]):%
\begin{equation}
\frac{1}{r^{2}}\approx \alpha ^{2}\left[ d_{0}+\frac{e^{-\alpha r}}{\left(
1-e^{-\alpha r}\right) ^{2}}\right] =\alpha ^{2}\left[ d_{0}+\frac{1}{\left(
\alpha r\right) ^{2}}-\frac{1}{12}+\frac{\left( \alpha r\right) ^{2}}{240}-%
\frac{\left( \alpha r\right) ^{4}}{6048}+\frac{\left( \alpha r\right) ^{6}}{%
172800}+O\left( \left( \alpha r\right) ^{8}\right) \right] .
\end{equation}%
When $\alpha r\ll 1,$ the value of the dimensionless constant $d_{0}=1/12$
has simply determined by the above series expansion and $\alpha $ takes the
unit of reciprocal of $r$. The present approximation was shown to be more
powerful than the usual approximation [40]. Obviously, the above
approximation to the centrifugal (pseudo-centrifugal) term turns to $1/r^{2}$
when the parameter $\alpha $ goes to zero as%
\begin{equation}
\underset{\alpha \rightarrow 0}{\lim }\left[ \alpha ^{2}\left( d_{0}+\frac{1%
}{e^{\alpha r}-1}+\frac{1}{\left( e^{\alpha r}-1\right) ^{2}}\right) \right]
=\frac{1}{r^{2}},
\end{equation}%
which shows that the usual approximation is the limit of our approximation
(cf. e.g., [17,40] and the references therein).

We introduce the following new dimensionless parameter, z$(r)=e^{-\alpha
r}\in \lbrack 0,1]$, which maintains the finiteness of the transformed wave
functions on the boundary conditions$.$ Thus$,$ substituting Eq. (15) into
Eq. (13), we obtain the following Schr\"{o}dinger-like equation satisfying $%
F_{n\kappa }(r),$%
\begin{equation*}
\left( \frac{d^{2}}{dz^{2}}+\frac{(1-z)}{z(1-z)}\frac{d}{dz}\right)
F_{n\kappa }(z)
\end{equation*}%
\begin{equation}
+\frac{1}{z^{2}(1-z)^{2}}\left\{ -\left[ \left( 2+b\right) b\nu
_{1}^{2}+\varepsilon _{n\kappa }^{2}\right] z^{2}+\left[ 2\left( b\nu
_{1}^{2}+\varepsilon _{n\kappa }^{2}\right) -\kappa \left( \kappa +1\right) %
\right] z-\varepsilon _{n\kappa }^{2}\right\} F_{n\kappa }(z)=0,
\end{equation}%
where 
\begin{equation}
\varepsilon _{n\kappa }^{2}=\nu _{1}^{2}-\omega _{1}^{2}+\kappa \left(
\kappa +1\right) d_{0},
\end{equation}%
and $F_{n\kappa }(r)=F_{n\kappa }(z).$ In order to clarify the parametric
generalization of the NU method [25], let us take the following general form
of a Schr\"{o}dinger-like equation written for any potential, 
\begin{equation}
\frac{d^{2}\psi _{n}}{dz^{2}}+\frac{\widetilde{\tau }(z)}{\sigma (z)}\frac{%
d\psi _{n}}{dz}+\frac{\widetilde{\sigma }(z)}{\sigma ^{2}(z)}\psi _{n}(z)=0,
\end{equation}%
satisfying the wave functions%
\begin{equation}
\psi _{n}(z)=\phi (z)y_{n}(z),
\end{equation}%
where 
\begin{equation}
\widetilde{\tau }(z)=c_{1}-c_{2}z,
\end{equation}%
and 
\begin{equation}
\sigma (z)=z\left( 1-c_{3}z\right) \text{ and \ }\widetilde{\sigma }%
(z)=-Az^{2}+Bz-C,
\end{equation}%
are two polynomials at most of first- and second-degree, respectively.
Furthermore, when Eq. (17) is compared with its counterpart Eq. (19), we can
obtain the specific values for the constants $c_{i}$ ($i=1,$2$,3)$ along
with $A,B$ and $C.$ Now, following the NU method [38] and making the
substitution of Eqs. (21) and (22), we can obtain general forms for the
polynomials $\pi (z)$ and $\tau (z),$ the root of the parameter $k,$ the
eigenvalues equation and the wave functions $\phi (z)$ and $y_{n}(z)$
expressed in terms of the constants $c_{i}$ ($i=4,$ $5,\cdots ,13)$ as shown
in Appendix A (cf. Refs. [18,21,25,40])$.$ Therefore, the task of computing
the energy eigenvalues and the corresponding wave functions of Eq. (17)
within the framework of the parametric generalization of the NU method is
relatively easy and straightforward. It is explained in the following steps.
Firstly, we need to find the specific values for the parametric constants $%
c_{i}$ ($i=4,$ $5,\cdots ,13)$ by means of the relation A1 of Appendix A.
The values of all these constants $c_{i}$ ($i=1,$ $2,\cdots ,13)$ together
with $A,$ $B$ and $C$ are therefore displayed in Table 1 for the GMP model.
Secondly, using the relations (A2-A5), the analytic forms of the essential
polynomials $\pi (z)$ and $\tau (z)$ along with the root $k,$ required by
the method [41], can also be found as%
\begin{equation}
\pi (z)=-\frac{z}{2}-\frac{1}{2}\left[ \left( \sqrt{\left( 1+2\kappa \right)
^{2}+4b^{2}\nu _{1}^{2}}+2\varepsilon _{n\kappa }\right) -2\varepsilon
_{n\kappa }\right] ,
\end{equation}%
\begin{equation}
k=2b\nu _{1}^{2}-\kappa \left( \kappa +1\right) -\varepsilon _{n\kappa }%
\sqrt{\left( 1+2\kappa \right) ^{2}+4b^{2}\nu _{1}^{2}},
\end{equation}%
\begin{equation}
\tau (z)=1+2\varepsilon _{n\kappa }-2\left( 1+\varepsilon _{n\kappa }+\frac{1%
}{2}\sqrt{\left( 1+2\kappa \right) ^{2}+4b^{2}\nu _{1}^{2}}\right) z,
\end{equation}%
where $\tau ^{\prime }(z)<0$ must be satisfied in order to obtain a physical
solution according to the NU method [38]$.$ Thirdly, we need to calculate
the energy eigenvalues by means of the eigenvalue equation, relation A6, and
obtain%
\begin{equation}
\varepsilon _{n\kappa }=\frac{\left( 2+b\right) b\nu _{1}^{2}}{2n+1+\sqrt{%
\left( 1+2\kappa \right) ^{2}+4b^{2}\nu _{1}^{2}}}-\frac{2n+1+\sqrt{\left(
1+2\kappa \right) ^{2}+4b^{2}\nu _{1}^{2}}}{4}.
\end{equation}%
Finally, with the aid of Eqs. (14) and (18), Eq. (26) can be also reduced to
the energy equation for the GMP with the spin symmetry concept for any
spin-orbit quantum number $\kappa =\pm 1,\pm 2,\cdots $ values in the Dirac
theory, 
\begin{equation*}
\left( Mc^{2}+E_{n\kappa }-C_{s}\right) \left( Mc^{2}+D-E_{n\kappa }\right)
+\hbar ^{2}c^{2}\alpha ^{2}\kappa \left( \kappa +1\right) d_{0}
\end{equation*}%
\begin{equation}
=\hbar ^{2}c^{2}\alpha ^{2}\left( \frac{\left( 2+b\right) b\nu _{1}^{2}}{%
2\left( n+\delta _{1}\right) }-\frac{\left( n+\delta _{1}\right) }{2}\right)
^{2},\text{ }n=0,1,2,\cdots ,
\end{equation}%
where%
\begin{equation}
\delta _{1}=\frac{1}{2}\left( 1+\sqrt{\left( 1+2\kappa \right)
^{2}+4b^{2}\nu _{1}^{2}}\right) \geq 1.
\end{equation}%
In what follows, in order to establish the wave functions $F_{n\kappa }(r)$
of Eq. (8), the relations (A7-A10) are used. Firstly, we find the first part
of the wave functions yields%
\begin{equation}
\phi (z)=z^{\varepsilon _{n\kappa }}(1-z)^{\delta _{1}},\text{ }\varepsilon
_{n\kappa }>0,\text{ }\delta _{1}>0.
\end{equation}%
Secondly, we calculate the weight function as%
\begin{equation}
\rho (z)=z^{2\varepsilon _{n\kappa }}(1-z)^{2\delta _{1}-1}.
\end{equation}%
and this, in turn, generates the second part of the wave functions,%
\begin{equation}
y_{n}(z)\sim z^{-2\varepsilon _{n\kappa }}(1-z)^{-\left( 2\delta
_{1}-1\right) }\frac{d^{n}}{dz^{n}}\left[ z^{n+2\varepsilon _{n\kappa
}}\left( 1-z\right) ^{n+2\delta _{1}-1}\right] \approx P_{n}^{(2\varepsilon
_{n\kappa },2\delta _{1}-1)}(1-2z),
\end{equation}%
where $P_{n}^{(a,b)}(1-2z)$ is the orthogonal Jacobi polynomials [49,50].
Finally, the upper spinor component $F_{n\kappa }(z)$ for arbitrary $\kappa $
can be obtained by means of Eq. (20) as%
\begin{equation*}
F_{n\kappa }(r)=\mathcal{N}_{n\kappa }e^{-\varepsilon _{n\kappa }\alpha
r}(1-e^{-\alpha r})^{\delta _{1}}P_{n}^{(2\varepsilon _{n\kappa },2\delta
_{1}-1)}(1-2e^{-\alpha r})
\end{equation*}%
\begin{equation}
=\mathcal{N}_{n\kappa }\frac{\Gamma (n+2\varepsilon _{n\kappa }+1)}{\Gamma
(2\varepsilon _{n\kappa }+1)n!}e^{-\varepsilon _{n\kappa }\alpha
r}(1-e^{-\alpha r})^{\delta _{1}}%
\begin{array}{c}
_{2}F_{1}%
\end{array}%
\left( -n,n+2\varepsilon _{n\kappa }+2\delta _{1};1+2\varepsilon _{n\kappa
};e^{-\alpha r}\right) ,
\end{equation}%
where the normalization constants $\mathcal{N}_{n\kappa }$ are calculated in
Appendix B.

Let us recall the derivative relation of the hypergeometric function, 
\begin{equation*}
\frac{d}{dz}\left[ 
\begin{array}{c}
_{2}F_{1}%
\end{array}%
\left( a;b;c;z\right) \right] =\left( \frac{ab}{c}\right) 
\begin{array}{c}
_{2}F_{1}%
\end{array}%
\left( a+1;b+1;c+1;z\right) ,
\end{equation*}%
that is used to calculate the corresponding lower-component $G_{n\kappa }(r)$
in Eq. (9) as 
\begin{equation*}
G_{n\kappa }(r)=\frac{\mathcal{N}_{n\kappa }}{Mc^{2}+E_{n\kappa }-C_{s}}%
\left( \frac{\alpha \delta _{1}e^{-\alpha r}}{1-e^{-\alpha r}}-\alpha
\varepsilon _{n\kappa }+\frac{\kappa }{r}\right) F_{n\kappa }(r)
\end{equation*}%
\begin{equation*}
+\mathcal{N}_{n\kappa }\frac{n\alpha \left( n+2\varepsilon _{n\kappa
}+2\delta _{1}\right) }{\left( Mc^{2}+E_{n\kappa }-C_{s}\right) \left(
1+2\varepsilon _{n\kappa }\right) }(1-e^{-\alpha r})^{\delta _{1}}\left(
e^{-\alpha r}\right) ^{\varepsilon _{n\kappa }+1}
\end{equation*}%
\begin{equation}
\times 
\begin{array}{c}
_{2}F_{1}%
\end{array}%
\left( 1-n;n+2\left( \varepsilon _{n\kappa }+\delta _{1}\right) +1;2\left(
1+\varepsilon _{n\kappa }\right) ;e^{-\alpha r}\right) .
\end{equation}%
We would like to note that the hypergeometric series $%
\begin{array}{c}
_{2}F_{1}%
\end{array}%
\left( 1-n;n+2\left( \varepsilon _{n\kappa }+\delta _{1}\right) +1;2\left(
1+\varepsilon _{n\kappa }\right) ;e^{-\alpha r}\right) $ terminates for $n=0$
and thus does not diverge for all values of real parameters $\varepsilon
_{n\kappa }$ and $\delta _{1}.$

In the presence of exact spin symmetry $\left( C_{s}=0\right) $, $E_{n\kappa
}\neq -Mc^{2}$ (only positive energy states do exist). In the exact spin
symmetry ($C_{s}=0$) with $(\kappa =1,\kappa =-2),$ the upper- and lower-
spinor wave functions for the ground $(0p_{1/2},0p_{3/2})$ and first excited 
$(1p_{1/2},1p_{3/2})$ degenerate eigenstates are being illustrated in Figure
2a and 2b, respectively. A glance at Figure 2 reveals that there is only one
curve (dashed line) for the radial wavefunctions of the upper components for
both states in the doublet. However, there are two curves (solid lines) for
the radial wavefunctions of the lower components. The following values of
the parameters$\ M=1.0$ $fm^{-1},$ $D=15$ $fm^{-1},$ $\alpha =0.1$ $fm^{-1}$
and $r_{e}=0.4$ $fm,$ $E_{0,\kappa =1}=E_{0,\kappa =-2}=5.5791076$ $fm^{-1}$
and $E_{1,\kappa =1}=E_{1,\kappa =-2}=8.1823677$ $fm^{-1}$ have been used.

Let us now study the special case when the parameter $\alpha \rightarrow 0$
in Eq. (1)$.$ The GMP potential can be easily reduced to the well-known
Kratzer molecular potential,%
\begin{equation}
\lim_{\alpha \rightarrow 0}V(r)=D\left( \frac{r-r_{e}}{r}\right) ^{2}.
\end{equation}%
which has been studied extensively by using different methods as the
function analysis [51], the NU [52,53] and the exact quantization rule (EQR)
[54]. To avoid the repetition, following Appendix A, we write down the
essential polynomials:%
\begin{equation}
\pi (r)=\frac{1}{2}\left[ 1+\gamma -2\epsilon _{n\kappa }r\right] ,
\end{equation}%
\begin{equation}
k=\frac{2r_{e}}{\hbar ^{2}c^{2}}\left( Mc^{2}+E_{n\kappa }-C_{s}\right)
D-\gamma \epsilon _{n\kappa },
\end{equation}%
\begin{equation}
\tau (r)=1+\gamma -2\epsilon _{n\kappa }r,
\end{equation}%
where%
\begin{equation}
\gamma =\sqrt{\left( 1+2\kappa \right) ^{2}+\frac{4r_{e}^{2}}{\hbar ^{2}c^{2}%
}\left( Mc^{2}+E_{n\kappa }-C_{s}\right) D},
\end{equation}%
\ \ 
\begin{equation}
\epsilon _{n\kappa }=\frac{1}{\hbar c}\sqrt{\left( Mc^{2}-E_{n\kappa
}+D\right) \left( Mc^{2}+E_{n\kappa }-C_{s}\right) }.
\end{equation}%
We further obtain the following two expressions which are relevant in the
construction of the energy equation (relation A6) [38]%
\begin{equation}
\lambda _{n}=2n\epsilon _{n\kappa }\text{ and }\lambda =\frac{2r_{e}}{\hbar
^{2}c^{2}}\left( Mc^{2}+E_{n\kappa }-C_{s}\right) D-(1+\gamma )\epsilon
_{n\kappa },
\end{equation}%
and substituting $\lambda _{n}=$ $\lambda ,$ we can obtain the following
spin symmetric energy equation, 
\begin{subequations}
\begin{equation}
\left( Mc^{2}-E_{n\kappa }+D\right) =\frac{qD^{2}\left( Mc^{2}+E_{n\kappa
}-C_{s}\right) }{\left( N_{n}+\sqrt{N_{\kappa }^{2}+qD\left(
Mc^{2}+E_{n\kappa }-C_{s}\right) }\right) ^{2}},
\end{equation}%
where $N_{n}=2n+1,$ $N_{\kappa }=2\kappa +1$ and $q=\left( 2r_{e}/\hbar
c\right) ^{2}.$ The above equation for energies looks like a quartic
equation of the form:%
\begin{equation}
a_{4}E_{n\kappa }^{4}+a_{3}E_{n\kappa }^{3}+a_{2}E_{n\kappa
}^{2}+a_{1}E_{n\kappa }+a_{0}=0
\end{equation}%
with coefficients 
\end{subequations}
\begin{equation*}
a_{4}=q^{2}D^{2};\text{ }a_{3}=2qD\left\lceil N_{\kappa
}^{2}-N_{n}^{2}-qDC_{s}\right\rceil ;
\end{equation*}%
\begin{equation*}
\text{ }a_{2}=(N_{n}^{2}-N_{\kappa
}^{2})^{2}+2qD(D+M+C_{s})(N_{n}^{2}-N_{\kappa
}^{2})+4qD^{2}N_{n}^{2}+q^{2}D^{2}\left( C_{s}^{2}+2MC_{s}-2M^{2}\right) ;
\end{equation*}%
\begin{eqnarray*}
a_{1} &=&2q^{2}D^{2}MC_{s}(M-C_{s})-2(D+M)(N_{n}^{2}-N_{\kappa }^{2})^{2} \\
&&+2qD\left( M^{2}-2MC_{s}-DC_{s}\right) (N_{n}^{2}-N_{\kappa
}^{2})-4qD^{2}\left( D+C_{s}\right) N_{n}^{2};
\end{eqnarray*}%
\begin{eqnarray*}
a_{0} &=&\left\lceil qDM^{2}-(D+M)(N_{n}^{2}-N_{\kappa }^{2})\right\rceil
^{2}-4qD^{2}\left( M-C_{s}\right) (D+M)N_{n}^{2} \\
&&+q^{2}D^{2}M^{2}C_{s}\left( C_{s}-2M\right)
+2qDMC_{s}(D+M)(N_{n}^{2}-N_{\kappa }^{2});
\end{eqnarray*}%
where we have set $c=1.$ For a given value of $n$ and $\kappa $ (or $l$),
the above quartic equation, Eq. (41b), provides four distinct positive and
negative real (real and complex) energy spectra related with $E_{n\kappa
}^{+}$ or $E_{n\kappa }^{-}$, respectively. One of the distinct solutions is
only valid to obtain the positive-energy bound states in the limit of the
spin symmetry. Therefore, the procedures for calculating the four distinct
energies; namely, $E_{n\kappa }^{(1)},$ $E_{n\kappa }^{(2)},$ $E_{n\kappa
}^{(3)}$ and $E_{n\kappa }^{(4)}$ are given in Appendix B$.$

Furthermore, following [18], the normalized upper- and lower-spinor wave
functions can be calculated as 
\begin{equation}
F_{n\kappa }(r)=\left( 2\epsilon _{n\kappa }\right) ^{K+1}\sqrt{\frac{%
\epsilon _{n\kappa }n!}{\left( n+K+1\right) \Gamma \left( n+2K+2\right) }}%
r^{K+1}e^{-\epsilon _{n\kappa }r}L_{n}^{\left( 2K+1\right) }(2\epsilon
_{n\kappa }r),
\end{equation}%
and%
\begin{equation*}
G_{n\kappa }(r)=\frac{1}{Mc^{2}+E_{n\kappa }-C_{s}}\left( 2\epsilon
_{n\kappa }\right) ^{K+1}\sqrt{\frac{\epsilon _{n\kappa }n!}{\left(
n+K+1\right) \Gamma \left( n+2K+2\right) }}
\end{equation*}%
\begin{equation*}
\times \left[ \left( \frac{\left( K+1\right) +\kappa }{r}-\epsilon _{n\kappa
}\right) F_{n\kappa }(r)-2\epsilon _{n\kappa }r^{K+1}e^{-\epsilon _{n\kappa
}r}L_{n-1}^{\left( 2K+2\right) }(2\epsilon _{n\kappa }r)\right] ,
\end{equation*}%
\begin{equation}
K=\frac{1}{2}\left( \gamma -1\right) ,
\end{equation}%
respectively, where $L_{n}^{\left( \beta \right) }(x)$ are associated
Laguerre polynomials. The simplest exact spin solution, representing the
ground state and first excited state, are%
\begin{equation*}
\left\{ 
\begin{array}{c}
\left( Mc^{2}+E_{0\kappa }\right) \left( Mc^{2}-E_{0\kappa }+D\right) =\frac{%
1}{\hbar ^{2}c^{2}}\left( \frac{r_{e}\left( Mc^{2}+E_{0\kappa }\right) D}{1+%
\sqrt{\left( 1+2\kappa \right) ^{2}+\frac{4Dr_{e}^{2}}{\hbar ^{2}c^{2}}%
\left( Mc^{2}+E_{0\kappa }\right) }}\right) ^{2}, \\ 
F_{0\kappa }(r)=\left( 2\epsilon _{0\kappa }\right) ^{K_{0}+1}\sqrt{\frac{%
\epsilon _{0\kappa }}{\left( K_{0}+1\right) \Gamma \left( 2K_{0}+2\right) }}%
r^{K_{0}+1}e^{-\epsilon _{n\kappa }r}, \\ 
G_{0\kappa }(r)=\frac{1}{Mc^{2}+E_{0\kappa }}\left( 2\epsilon _{0\kappa
}\right) ^{K_{0}+1}\sqrt{\frac{\epsilon _{0\kappa }}{\left( K_{0}+1\right)
\Gamma \left( 2K_{0}+2\right) }}\left( \frac{\left( K_{0}+1\right) +\kappa }{%
r}-\epsilon _{0\kappa }\right) F_{0\kappa }(r),%
\end{array}%
\right. \text{ }
\end{equation*}%
\begin{equation*}
\left\{ 
\begin{array}{c}
\left( Mc^{2}+E_{1\kappa }\right) \left( D+Mc^{2}-E_{1\kappa }\right) =\frac{%
1}{\hbar ^{2}c^{2}}\left( \frac{Dr_{e}\left( Mc^{2}+E_{1\kappa }\right) }{3+%
\sqrt{\left( 1+2\kappa \right) ^{2}+\frac{4Dr_{e}^{2}}{\hbar ^{2}c^{2}}%
\left( Mc^{2}+E_{1\kappa }\right) }}\right) ^{2}, \\ 
F_{1\kappa }(r)=\left( 2\epsilon _{1\kappa }\right) ^{K_{1}+1}\sqrt{\frac{%
\epsilon _{1\kappa }}{\left( K_{1}+2\right) \Gamma \left( 2K_{1}+3\right) }}%
r^{K_{1}+1}e^{-\epsilon _{1\kappa }r}\left( -2\epsilon _{1\kappa
}r+2K_{1}+2\right) , \\ 
G_{1\kappa }(r)=\frac{1}{Mc^{2}+E_{1\kappa }}\left( 2\epsilon _{1\kappa
}\right) ^{K_{1}+1}\sqrt{\frac{\epsilon _{1\kappa }}{\left( K+2\right)
\Gamma \left( 2K+3\right) }}\left[ \left( \frac{\left( K_{1}+1\right)
+\kappa }{r}-\epsilon _{1\kappa }\right) F_{1\kappa }(r)-2\epsilon _{1\kappa
}r^{K_{1}+1}e^{-\epsilon _{1\kappa }r}\right] ,%
\end{array}%
\right. \text{ }
\end{equation*}%
respectively. Finally, we would like to note that $\epsilon _{0\kappa }$ and 
$\epsilon _{1\kappa }$ can be calculated via Eq. (39) whereas $K_{0}$ and $%
K_{1}$ via Eq. (43) along with Eq. (38) when $C_{s}=0.$

\subsection{Pseudospin symmetry solutions of the GMP model}

From Eq. (10), we can see that the energy eigenvalues, $E_{n\kappa },$
depend only on $n$ and $\widetilde{l},$ i.e., $E_{n\kappa }=E(n,\widetilde{l}%
(\widetilde{l}+1)).$ For $\widetilde{l}\neq 0,$ the states with $j=%
\widetilde{l}\pm 1/2$ are degenerate. This is a $SU(2)$ pseudospin symmetry.
Following Refs. [22-25], we impose the GMP [1] as the $\Delta (r),$ i.e.,%
\begin{equation}
\Delta (r)=D\left( 1-\frac{b}{e^{\alpha r}-1}\right) ^{2},
\end{equation}%
leads us to obtain a Schr\"{o}dinger-like equation in the form:%
\begin{equation}
\left[ \frac{d^{2}}{dr^{2}}-\frac{\kappa \left( \kappa -1\right) }{r^{2}}%
+\alpha ^{2}\nu _{2}^{2}\left( 1-\frac{be^{-\alpha r}}{1-e^{-\alpha r}}%
\right) ^{2}+\alpha ^{2}\omega _{2}^{2}\right] G_{n\kappa }(r)=0
\end{equation}%
where%
\begin{equation}
\nu _{2}^{2}=\frac{1}{\alpha ^{2}\hbar ^{2}c^{2}}\left( Mc^{2}-E_{n\kappa
}+C_{ps}\right) D,\text{ \ }\omega _{2}^{2}=\frac{1}{\alpha ^{2}\hbar
^{2}c^{2}}\left[ E_{n\kappa }^{2}-M^{2}c^{4}-C_{ps}\left( Mc^{2}+E_{n\kappa
}\right) \right]
\end{equation}%
where $\kappa \left( \kappa -1\right) =\widetilde{l}(\widetilde{l}+1).$ We
follow the same procedures in the previous subsection to obtain a Dirac
equation satisfying $G_{n\kappa }(r),$%
\begin{equation*}
\left( \frac{d^{2}}{dz^{2}}+\frac{(1-z)}{z(1-z)}\frac{d}{dz}\right)
G_{n\kappa }(z)
\end{equation*}%
\begin{equation}
+\frac{1}{z^{2}(1-z)^{2}}\left\{ -\left[ \widetilde{\varepsilon }_{n\kappa
}^{2}-\left( 2+b\right) b\nu _{2}^{2}\right] z^{2}+\left[ 2\left( \widetilde{%
\varepsilon }_{n\kappa }^{2}-b\nu _{2}^{2}\right) -\kappa \left( \kappa
-1\right) \right] z-\widetilde{\varepsilon }_{n\kappa }^{2}\right\}
G_{n\kappa }(z)=0,
\end{equation}%
where 
\begin{equation}
\widetilde{\varepsilon }_{n\kappa }^{2}=-\omega _{2}^{2}-\nu _{2}^{2}+\kappa
\left( \kappa -1\right) d_{0},
\end{equation}%
where $G_{n\kappa }(r)=G_{n\kappa }(z).$ To avoid repetition in the solution
of Eq. (45), a careful inspection for the relationship between the present
set of parameters $(\widetilde{\varepsilon }_{n\kappa },\nu _{2}^{2})$ and
the previous one $(\varepsilon _{n\kappa },\nu _{1}^{2})$ tells us that the
negative energy solution for pseudospin symmetry, where $S(r)\sim -V(r),$
can be obtained directly from the spin symmetric solution by using the
following parameter mapping [18,55]: 
\begin{equation}
F_{n\kappa }(r)\leftrightarrow G_{n\kappa }(r),\kappa \rightarrow \kappa -1,%
\text{ }V(r)\rightarrow -V(r)\text{ (i.e., }D\rightarrow -D\text{)},\text{ }%
E_{n\kappa }\rightarrow -E_{n\kappa }\text{ and }C_{s}\rightarrow -C_{ps}.
\end{equation}%
Following the previous procedures, the constants in the case of pseudospin
symmetry concept are displayed in Table 2. Applying the above
transformations to Eqs. (27), (32) and (33) lead to the following pseudospin
symmetric energy equation,%
\begin{equation*}
\left( E_{n\kappa }-Mc^{2}-C_{ps}\right) \left( D-Mc^{2}-E_{n\kappa }\right)
+\hbar ^{2}c^{2}\alpha ^{2}\kappa \left( \kappa -1\right) d_{0}
\end{equation*}%
\begin{equation}
=\hbar ^{2}c^{2}\alpha ^{2}\left( \frac{\left( 2+b\right) b\nu _{2}^{2}}{%
2\left( n+\delta _{2}\right) }+\frac{\left( n+\delta _{2}\right) }{2}\right)
^{2},\text{ }n=0,1,2,\cdots ,
\end{equation}%
with%
\begin{equation}
\delta _{2}=\frac{1}{2}\left( 1+\sqrt{\left( 1-2\kappa \right)
^{2}-4b^{2}\nu _{2}^{2}}\right) \geq 1.
\end{equation}%
Note that Eq. (50) can be also expressed in the form of quartic equation
(cf. Eq. (41b)) since the two parameters $\nu _{2}^{2}$ and $\delta _{2}$
contain the energy eigenvalues $E_{n\kappa }.$ The procedures of this
analytic solution is so similar to the one presented in Appendix B with the
changes $D\rightarrow -D,$ $E_{n\kappa }\rightarrow -E_{n\kappa }$ and $%
C_{s}\rightarrow -C_{ps}.$ Furthermore, the lower-component wave functions: 
\begin{equation*}
G_{n\kappa }(r)=\mathcal{N}_{n\kappa }e^{-\widetilde{\varepsilon }_{n\kappa
}\alpha r}(1-e^{-\alpha r})^{\delta _{2}}P_{n}^{(2\widetilde{\varepsilon }%
_{n\kappa },2\delta _{2}-1)}(1-2e^{-\alpha r})
\end{equation*}%
\begin{equation}
=\mathcal{N}_{n\kappa }\frac{\left( 2\widetilde{\varepsilon }_{n\kappa
}+1\right) _{n}}{n!}e^{-\widetilde{\varepsilon }_{n\kappa }\alpha
r}(1-e^{-\alpha r})^{\delta _{2}}%
\begin{array}{c}
_{2}F_{1}%
\end{array}%
\left( -n,n+2\widetilde{\varepsilon }_{n\kappa }+2\delta _{2};1+2\widetilde{%
\varepsilon }_{n\kappa };e^{-\alpha r}\right) ,
\end{equation}%
where%
\begin{equation}
\widetilde{\varepsilon }_{n\kappa }=-\left( \frac{\left( 2+b\right) b\nu
_{2}^{2}}{2n+1+\sqrt{\left( 1-2\kappa \right) ^{2}-4b^{2}\nu _{2}^{2}}}+%
\frac{2n+1+\sqrt{\left( 1-2\kappa \right) ^{2}-4b^{2}\nu _{2}^{2}}}{4}%
\right) ,
\end{equation}%
The upper-component $F_{n\kappa }(r)$ can be calculated from Eq. (11) as
follows 
\begin{equation*}
F_{n\kappa }(r)=\frac{\mathcal{N}_{n\kappa }}{Mc^{2}-E_{n\kappa }+C_{ps}}%
\left( \frac{\alpha \delta _{2}e^{-\alpha r}}{1-e^{-\alpha r}}-\alpha 
\widetilde{\varepsilon }_{n\kappa }-\frac{\kappa }{r}\right) G_{n\kappa }(r)
\end{equation*}%
\begin{equation*}
+\mathcal{N}_{n\kappa }\frac{n\alpha \left( n+2\widetilde{\varepsilon }%
_{n\kappa }+2\delta _{2}\right) }{\left( Mc^{2}-E_{n\kappa }+C_{ps}\right)
\left( 1+2\widetilde{\varepsilon }_{n\kappa }\right) }(1-e^{-\alpha
r})^{\delta _{2}}\left( e^{-\alpha r}\right) ^{\widetilde{\varepsilon }%
_{n\kappa }+1}
\end{equation*}%
\begin{equation}
\times 
\begin{array}{c}
_{2}F_{1}%
\end{array}%
\left( 1-n;n+2\left( \widetilde{\varepsilon }_{n\kappa }+\delta _{2}\right)
+1;2\left( 1+\widetilde{\varepsilon }_{n\kappa }\right) ;e^{-\alpha
r}\right) .
\end{equation}%
Hence, in the exact pseudospin symmetry where $C_{ps}=0$, $E_{n\kappa }\neq
Mc^{2}$ (only negative energy states exist).

On the other hand, the pseudospin solutions of the Dirac equation for the
Kratzer potential can be obtained from the spin symmetry case by applying
transformation map given by Eq. (49) as%
\begin{equation}
\widetilde{\gamma }=\sqrt{\left( 1-2\kappa \right) ^{2}-\frac{4Dr_{e}^{2}}{%
\hbar ^{2}c^{2}}\left( Mc^{2}-E_{n\kappa }+C_{ps}\right) },
\end{equation}%
\begin{equation}
\widetilde{\epsilon }_{n\kappa }=\frac{1}{\hbar c}\sqrt{-D\left(
Mc^{2}-E_{n\kappa }+C_{ps}\right) -\left( E_{n\kappa
}^{2}-M^{2}c^{4}-C_{ps}\left( Mc^{2}+E_{n\kappa }\right) \right) }.
\end{equation}%
Therefore, the eigenvalue equation is%
\begin{equation*}
E_{n\kappa }^{2}-M^{2}c^{4}-C_{ps}\left( Mc^{2}+E_{n\kappa }\right)
=-D\left( Mc^{2}-E_{n\kappa }+C_{ps}\right)
\end{equation*}%
\begin{equation}
-\frac{1}{\hbar ^{2}c^{2}}\left( \frac{Dr_{e}\left( Mc^{2}-E_{n\kappa
}+C_{ps}\right) }{n+K+1}\right) ^{2},\text{ }\widetilde{K}=\frac{1}{2}\left( 
\widetilde{\gamma }-1\right) ,
\end{equation}%
and the normalized lower- and upper-spinor wave functions are given by%
\begin{equation}
G_{n\kappa }(r)=\left( 2\widetilde{\epsilon }_{n\kappa }\right) ^{\widetilde{%
K}+1}\sqrt{\frac{\widetilde{\epsilon }_{n\kappa }n!}{\left( n+\widetilde{K}%
+1\right) \Gamma \left( n+2\widetilde{K}+2\right) }}r^{\widetilde{K}+1}e^{-%
\widetilde{\epsilon }_{n\kappa }r}L_{n}^{\left( 2\widetilde{K}+1\right) }(2%
\widetilde{\epsilon }_{n\kappa }r),
\end{equation}%
\begin{equation*}
F_{n\kappa }(r)=\frac{1}{Mc^{2}-E_{n\kappa }+C_{ps}}\left( 2\widetilde{%
\epsilon }_{n\kappa }\right) ^{\widetilde{K}+1}\sqrt{\frac{\widetilde{%
\epsilon }_{n\kappa }n!}{\left( n+\widetilde{K}+1\right) \Gamma \left( n+2%
\widetilde{K}+2\right) }}
\end{equation*}%
\begin{equation}
\times \left[ \left( \frac{\left( \widetilde{K}+1\right) -\kappa }{r}-%
\widetilde{\epsilon }_{n\kappa }\right) G_{n\kappa }(r)-2\widetilde{\epsilon 
}_{n\kappa }r^{\widetilde{K}+1}e^{-\widetilde{\epsilon }_{n\kappa
}r}L_{n-1}^{\left( 2\widetilde{K}+2\right) }(2\widetilde{\epsilon }_{n\kappa
}r)\right] .
\end{equation}

\section{Some Special Cases}

Let us study three special cases. At first, we study the $s(\widetilde{s})$%
-states ($l=\widetilde{l}=0,$ i.e., $\kappa =\mp 1$ ). It follows that the
spin-orbit coupling term $\kappa (\kappa +1)/r^{2}=0,$ and also the
corresponding approximation to it in Eq. (27). In the presence of the exact
spin symmetry limit $(C_{s}=0)$, the $s$-states ($\kappa =-1$) energy
equation becomes%
\begin{equation}
E_{n,-1}^{2}-M^{2}c^{4}-D\left( Mc^{2}+E_{n,-1}\right) =-\hbar
^{2}c^{2}\alpha ^{2}\left( \frac{\frac{1}{\hbar ^{2}c^{2}\alpha ^{2}}\left(
2+b\right) \left( Mc^{2}+E_{n,-1}\right) Db}{2\left( n+\delta \right) }-%
\frac{n+\delta }{2}\right) ^{2},
\end{equation}%
where%
\begin{equation}
\delta =\frac{1}{2}\left( 1+\sqrt{1+\frac{4}{\hbar ^{2}c^{2}\alpha ^{2}}%
\left( Mc^{2}+E_{n,-1}\right) Db^{2}}\right) \geq 1.
\end{equation}%
The upper- and lower-spinor wave functions are%
\begin{equation}
F_{n,-1}(r)=\mathcal{N}_{n,-1}\frac{\left( 2\eta +1\right) _{n}}{n!}e^{-\eta
\alpha r}(1-e^{-\alpha r})^{\delta }%
\begin{array}{c}
_{2}F_{1}%
\end{array}%
\left( -n,n+2\eta +2\delta ;1+2\eta ;e^{-\alpha r}\right) ,
\end{equation}%
and 
\begin{equation*}
G_{n,-1}(r)=\frac{\mathcal{N}_{n,-1}}{Mc^{2}+E_{n,-1}}\left( \frac{\alpha
\delta e^{-\alpha r}}{1-e^{-\alpha r}}-\alpha \eta -\frac{1}{r}\right)
F_{n,-1}(r)
\end{equation*}%
\begin{equation*}
+\mathcal{N}_{n,-1}\frac{n\alpha \left( n+2\eta +2\delta \right) }{\left(
Mc^{2}+E_{n,-1}\right) \left( 1+2\eta \right) }(1-e^{-\alpha r})^{\delta
}\left( e^{-\alpha r}\right) ^{\eta +1}
\end{equation*}%
\begin{equation}
\times 
\begin{array}{c}
_{2}F_{1}%
\end{array}%
\left( 1-n;n+2\left( \eta +\delta \right) +1;2\left( 1+\eta \right)
;e^{-\alpha r}\right) ,
\end{equation}%
with%
\begin{equation}
\eta =\frac{\frac{1}{\hbar ^{2}c^{2}\alpha ^{2}}\left( 2+b\right) \left(
Mc^{2}+E_{n,-1}\right) Db}{2\left( n+\delta \right) }-\frac{n+\delta }{2},
\end{equation}%
where $\mathcal{N}_{n,-1}$ is calculated in the Appendix C. Overmore, for
the $\widetilde{s}$-states ($\kappa =1$) in the exact pseudospin symmetry $%
(C_{ps}=0)$, the energy equation (48) becomes%
\begin{equation}
E_{n1}^{2}-M^{2}c^{4}+D\left( Mc^{2}-E_{n1}\right) =-\hbar ^{2}c^{2}\alpha
^{2}\left( \frac{\frac{1}{\hbar ^{2}c^{2}\alpha ^{2}}\left( 2+b\right)
\left( Mc^{2}-E_{n1}\right) Db}{2\left( n+\delta _{2}\right) }+\frac{\left(
n+\delta _{2}\right) }{2}\right) ^{2},
\end{equation}%
with%
\begin{equation}
\delta _{2}=\frac{1}{2}\left( 1+\sqrt{1-\frac{4}{\alpha ^{2}\hbar ^{2}c^{2}}%
\left( mc^{2}-E_{n1}\right) Db^{2}}\right) ,\text{ }E_{n1}>mc^{2},\text{ }%
n=0,1,2,\cdots .
\end{equation}%
The wave functions given by Eqs. (52) and (54) become%
\begin{equation*}
G_{n1}(r)=\mathcal{N}_{n1}e^{-\eta _{2}\alpha r}(1-e^{-\alpha r})^{\delta
_{2}}P_{n}^{(2\eta _{2},2\delta _{2}-1)}(1-2e^{-\alpha r})
\end{equation*}%
\begin{equation}
=\mathcal{N}_{n1}\frac{\left( 2\eta _{2}+1\right) _{n}}{n!}e^{-\eta
_{2}\alpha r}(1-e^{-\alpha r})^{\delta _{2}}%
\begin{array}{c}
_{2}F_{1}%
\end{array}%
\left( -n,n+2\eta _{2}+2\delta _{2};1+2\eta _{2};e^{-\alpha r}\right) ,
\end{equation}%
with%
\begin{equation}
\eta _{2}=\frac{\left( 2+b\right) \xi _{2}}{2n+1+\sqrt{1+4b\xi _{2}}}-\frac{%
2n+1+\sqrt{1+4b\xi _{2}}}{4},
\end{equation}%
and 
\begin{equation*}
F_{n1}(r)=\frac{\mathcal{N}_{n1}}{Mc^{2}-E_{n1}}\left( \frac{\alpha \delta
_{2}e^{-\alpha r}}{1-e^{-\alpha r}}-\alpha \eta _{2}-\frac{1}{r}\right)
G_{n\kappa }(r)
\end{equation*}%
\begin{equation*}
+\mathcal{N}_{n1}\frac{n\alpha \left( n+2\eta _{2}+2\delta _{2}\right) }{%
\left( Mc^{2}-E_{n\kappa }\right) \left( 1+2\eta _{2}\right) }(1-e^{-\alpha
r})^{\delta _{2}}\left( e^{-\alpha r}\right) ^{\eta _{2}+1}
\end{equation*}%
\begin{equation}
\times 
\begin{array}{c}
_{2}F_{1}%
\end{array}%
\left( 1-n;n+2\left( \eta _{2}+\delta _{2}\right) +1;2\left( 1+\eta
_{2}\right) ;e^{-\alpha r}\right) ,
\end{equation}%
where $E_{n1}\neq Mc^{2}.$

Second, we study the nonrelativistic limit. In applying the following
appropriate mapping \ $E_{n\kappa }-Mc^{2}\rightarrow E_{nl}$ and $\frac{1}{%
\hbar ^{2}c^{2}}\left( Mc^{2}+E_{n\kappa }\right) \rightarrow \frac{2\mu }{%
\hbar ^{2}}$ to Eqs. (27) and (32)$,$ we obtain the energy levels of the Schr%
\"{o}dinger equation for any arbitrary orbital quantum number $l$ as%
\begin{equation}
E_{nl}=D+\frac{\hbar ^{2}}{2\mu }l\left( l+1\right) \alpha ^{2}d_{0}-\frac{%
\hbar ^{2}\alpha ^{2}}{2\mu }\left( \frac{\frac{\mu }{\hbar ^{2}\alpha ^{2}}%
\left( 2+b\right) Db}{n+\widetilde{\delta }}-\frac{n+\widetilde{\delta }}{2}%
\right) ^{2},
\end{equation}%
and the radial wave functions as%
\begin{equation}
\psi _{nl}(r)=\mathcal{N}_{nl}\frac{\left( 2\widetilde{\eta }+1\right) _{n}}{%
n!}e^{-\widetilde{\eta }\alpha r}(1-e^{-\alpha r})^{\widetilde{\delta }}%
\begin{array}{c}
_{2}F_{1}%
\end{array}%
\left( -n,n+2\widetilde{\eta }+2\widetilde{\delta };1+2\widetilde{\eta }%
;e^{-\alpha r}\right) ,
\end{equation}%
with%
\begin{equation}
\widetilde{\delta }=\frac{1}{2}\left( 1+\sqrt{\left( 1+2l\right) ^{2}+\frac{%
8\mu }{\hbar ^{2}\alpha ^{2}}Db^{2}}\right) \geq 1,\text{ }\widetilde{\eta }=%
\frac{\frac{\mu }{\hbar ^{2}\alpha ^{2}}\left( 2+b\right) Db}{n+\widetilde{%
\delta }}-\frac{n+\widetilde{\delta }}{2},
\end{equation}%
where $\mu =m_{1}m_{2}/(m_{1}+m_{2})$ is the reduced mass of the two atoms
and $\mathcal{N}_{nl}$ is calculated in the Appendix C.

Third, the case $\alpha \rightarrow 0,$ the nonrelativistic limit of the
bound state solutions of Eqs. (41) and (42) in exact symmetry limit ($%
C_{s}=0)$ can be obtained as 
\begin{equation}
E_{nl}=D--\frac{8\mu }{\hbar ^{2}}\left( \frac{Dr_{e}}{1+2n+\sqrt{\left(
1+2l\right) ^{2}+\frac{8\mu }{\hbar ^{2}}Dr_{e}^{2}}}\right) ^{2},
\end{equation}%
and the normalized wave functions: 
\begin{equation}
\psi _{nl}(r)=\left( 2\widetilde{\epsilon }_{nl}\right) ^{L+1}\sqrt{\frac{%
\widetilde{\epsilon }_{nl}n!}{\left( n+L+1\right) \Gamma \left(
n+2L+2\right) }}r^{L+1}e^{-\widetilde{\epsilon }_{nl}r}L_{n}^{\left(
2L+1\right) }(2\widetilde{\epsilon }_{nl}r),
\end{equation}%
where%
\begin{equation}
\widetilde{\epsilon }_{nl}=\frac{2\mu Dr_{e}}{\hbar ^{2}(n+L+1)}\text{ \ and
\ \ }L=\frac{1}{2}\left( \sqrt{\left( 1+2l\right) ^{2}+\frac{8\mu }{\hbar
^{2}}Dr_{e}^{2}}-1\right) ,
\end{equation}%
which are identical to the previous results obtained by the function
analysis method [51], the NU method [52,53] and the EQR (cf. Eq. (29) in
Ref. [54] obtained in $D$-dimensions).

\section{Numerical results}

As in Ref. [56], we calculate the non-relativistic energy levels as function
of various values of the parameter that controls the width of the potential
well $\alpha =0.05$--$0.30$ $fm^{-1}$ and equilibrium inter-nuclear distance 
$r_{e}=0.4,$ $0.8$ $fm$ for various states with quantum numbers $n$ and $l.$
The atomic units $\hbar =M=1$ are used and the dissociation energy is set to 
$D=15$ $fm^{-1}.$ In Table 3, we display our results with those ones
calculated by using the conventional approximation scheme suggested by
Greene and Aldrich [12] to deal with the centrifugal term $l(l+1)/r^{2}$
together with the values obtained from the numerical integration procedures
based on the MATHEMATICA package programmed by Lucha and Sch\"{o}berl [57].
Obviously, our results are closely approaching the ones obtained in [57] for
both short range (small $\alpha $) as well as for long range (large $\alpha $%
) potential (see e.g., our recent works [39,58]). This means that our new
approximated calculations using the approximation scheme (15) proposed
recently by us provides much better approximation to the centrifugal term
than that in Ref. [55] even for large $\alpha $ values (see, e.g., [40] and
the references therein). At small values of $\alpha $ (Kratzer potential),
we have also calculated the energy levels as a function of $r_{e}=0.1-1.5$ $%
fm$ for various quantum numbers $n$ and $l.$ Hence, our numerical values of
these energy levels are shown in Table 4. Overmore, Table 5 presents some
numerical values for the eigeenergies of the Dirac valence states obtained
from Eq. (27) with parameters $M=1.0$ $fm^{-1},$ $D=15$ $fm^{-1},$ $\alpha
=0.1,0.3$ $fm^{-1}$ and $r_{e}=0.4,0.8$ $fm$ (exact spin symmetry case,
i.e., $C_{s}=0$ $fm^{-1}$)$.$ We noticed that there are only positive energy
bound state solutions in the spin symmetry limit. One can also see from
Table 5 that there are degeneracies between the eigenstates $%
(np_{1/2},np_{3/2}),$ $(nd_{3/2},nd_{5/2}),$ $(nf_{5/2},nf_{7/2}),$ $%
(ng_{7/2},ng_{9/2}),$ etc. In fact, each of these eigenstates form a spin
doublet. For instance, for any specific value of $n,$ where $n=0,1,2,\cdots
, $ $np_{1/2}$ with $\kappa =1$ is considered as the partner of $np_{3/2}$
with $\kappa =-2.$ Thus, states that have the same radial $n$ and orbital
angular momentum $l$ quantum numbers with \ $j=$ $l+1/2$ and $j=$ $l-1/2$
are degenerate [59].

On the other hand, in Table 6, we present some numerical values for the
eigeenergies of the Dirac hole states obtained from Eq. (50) with the
previous choice of potential parameters for the case of pseudospin symmetry
limit ($C_{ps}=0,$ $5.0,$ $-5.0$ and $-10.0$ $fm^{-1}$)$.$ One can see from
Table 6 that there are degeneracies between the eigenstates $%
(ns_{1/2},\left( n-1\right) d_{3/2}),$ $(np_{3/2},\left( n-1\right)
f_{5/2}), $ $(nd_{5/2},\left( n-1\right) g_{7/2}),$ $(nf_{7/2},\left(
n-1\right) h_{9/2}),$ etc. In fact, each of these eigenstates form a
pseudospin doublet. For instance, for specific value of $n=1,$ $1s_{1/2}$
with $\kappa =-1$ is considered as the partner of $0d_{3/2}$ with $\kappa
=2. $ Thus, states that have pseudo orbital angular momentum $\widetilde{l}$
quantum numbers, radial $n$ and $n-1$ with \ $j=\widetilde{l}-1/2$ and $\ j=$
$\widetilde{l}+1/2,$ respectively, are degenerate [59].

\section{Conclusions}

We have studied the approximate bound state solutions of the Dirac equation
for the GMP model with any arbitrary spin-orbit $\kappa $ state under the
conditions of the spin (pseudospin) symmetry limitation by means of the NU
method including a new improved approximation scheme to approximate the
centrifugal (pseudo-centrifugal) barrier term. By setting $\Sigma (r)$ ($%
\Delta (r)$) to the spherically symmetric GMP, we have derived the solutions
of the Dirac equation for the relativistic energy eigenvalues and associated
two-component spinor wave functions for arbitrary spin-orbit $\kappa $ state
that provides an approximate solution to the spin- and pseudo-spin symmetry.
The resulting solutions of the wave functions are being expressed in terms
of the Jacobi polynomials (or hypergeometric functions). We have shown that
the present spin symmetry can be easily reduced to the non-relativistic
solution once we set $V(r)=S(r)$ $($i.e., $\Delta (r)=0$ or $C_{s}=0$). The
non-relativistic limits of our solution are obtained by imposing appropriate
transformations and recalling $\kappa (\kappa +1)\rightarrow l(l+1)$ in the
spin symmetry limits. Furthermore, when $\alpha \rightarrow 0,$ our results
can be reduced to the well-known bound state solutions for the Kratzer
potential model. If we choose the spin-orbit quantum number $\kappa =-1$ ($%
\kappa =1$) for spin (pseudospin) symmetry$,$ the problem reduces to the
exact $s(\widetilde{s})$-wave Dirac solution. We must point out that the
numerical calculations for eigenenergy of the Dirac states involved in Eqs.
(27) and (50) are sensitive to the choice of the parameters $C_{s},$ $%
C_{ps}, $ $\alpha ,$ $r_{e},$ $D$ and $M.$ It is found that the spin
(pseudospin) limit Dirac eigenenergy valence (hole) states along with the
two-component spinors are identical with the results obtained previously by
other methods and works [15]. It is noticed that these results are obtained
in a much simpler fashion than Ref. [15]. Finally, Eqs. (27) and (50) can be
used to evaluate the binding energies of the GMP for diatomic molecules such
as $\mathrm{CH},$ $\mathrm{CO}$ and $\mathrm{N}_{2}$ [2,56] in the
relativistic framework with spin and pseudospin symmetry cases for any range
of potential well $\alpha $. Equations. (41) and (57) can be also used to
evaluate the binding energy of the Kratzer potential in the relativistic
framework with spin and pseudospin symmetry cases at small range potential
well ($\alpha \rightarrow 0$).

\acknowledgments We thank the anonymous referee and editors for their
enlightening comments and suggestions which helped us to significantly
improve the manuscript.

\appendix

\section{Generalization of the NU method}

We briefly review the Nikiforov-Uvarov essential polynomials, root,
eigenvalues and wave functions (see Eqs. (5) - (11) of Ref. [40]) being
expressed in terms of the parameters $c_{i}$ ($i=1,$ $2,\cdots ,13)$
together with $A,$ $B$ and $C$:

(i) The relevant constants:%
\begin{equation*}
c_{4}=\frac{1}{2}\left( 1-c_{1}\right) ,\text{ }c_{5}=\frac{1}{2}\left(
c_{2}-2c_{3}\right) ,\text{ }c_{6}=c_{5}^{2}+A,
\end{equation*}%
\begin{equation*}
\text{ }c_{7}=2c_{4}c_{5}-B,\text{ }c_{8}=c_{4}^{2}+C,\text{ }%
c_{9}=c_{3}\left( c_{7}+c_{3}c_{8}\right) +c_{6},
\end{equation*}%
\begin{equation*}
c_{10}=c_{1}+2c_{4}+2\sqrt{c_{8}}-1>-1,\text{ }c_{11}=1-c_{1}-2c_{4}+\frac{2%
}{c_{3}}\sqrt{c_{9}}>-1,
\end{equation*}%
\begin{equation}
c_{12}=c_{4}+\sqrt{c_{8}}>0,\text{ }c_{13}=-c_{4}+\frac{1}{c_{3}}\left( 
\sqrt{c_{9}}-c_{5}\right) >0.
\end{equation}%
(ii) The essential polynomials: 
\begin{equation}
\pi (z)=c_{4}+c_{5}z-\left[ \left( \sqrt{c_{9}}+c_{3}\sqrt{c_{8}}\right) z-%
\sqrt{c_{8}}\right] ,
\end{equation}%
\begin{equation}
k=-\left( c_{7}+2c_{3}c_{8}\right) -2\sqrt{c_{8}c_{9}},
\end{equation}%
\begin{equation}
\tau (z)=1-\left( c_{2}-2c_{5}\right) z-2\left[ \left( \sqrt{c_{9}}+c_{3}%
\sqrt{c_{8}}\right) z-\sqrt{c_{8}}\right] ,
\end{equation}%
\begin{equation}
\tau ^{\prime }(z)=-2c_{3}-2\left( \sqrt{c_{9}}+c_{3}\sqrt{c_{8}}\right) <0.
\end{equation}%
(iii) The energy equation:%
\begin{equation}
\left( c_{2}-c_{3}\right) n+c_{3}n^{2}-\left( 2n+1\right) c_{5}+\left(
2n+1\right) \left( \sqrt{c_{9}}+c_{3}\sqrt{c_{8}}\right) +c_{7}+2c_{3}c_{8}+2%
\sqrt{c_{8}c_{9}}=0.
\end{equation}%
(iv) The wave functions:%
\begin{equation}
\rho (z)=z^{c_{10}}(1-c_{3}z)^{c_{11}},
\end{equation}%
\begin{equation}
\phi (z)=z^{c_{12}}(1-c_{3}z)^{c_{13}},\text{ }c_{12}>0,\text{ }c_{13}>0,
\end{equation}%
\begin{equation}
y_{n}(z)=P_{n}^{\left( c_{10},c_{11}\right) }(1-2c_{3}z),\text{ }c_{10}>-1,%
\text{ }c_{11}>-1,
\end{equation}%
\begin{equation}
F_{n\kappa }(z)=\mathcal{N}_{n\kappa
}z^{c_{12}}(1-c_{3}z)^{c_{13}}P_{n}^{\left( c_{10},c_{11}\right)
}(1-2c_{3}z),
\end{equation}%
where $P_{n}^{\left( \mu ,\nu \right) }(x),$ $\mu >-1,\nu >-1$ and $x\in
\lbrack -1,1]$ are the Jacobi polynomials with%
\begin{equation}
P_{n}^{\left( \alpha ,\beta \right) }(1-2s)=\frac{\left( \alpha +1\right)
_{n}}{n!}%
\begin{array}{c}
_{2}F_{1}%
\end{array}%
\left( -n,1+\alpha +\beta +n;\alpha +1;s\right) ,
\end{equation}%
and $\mathcal{N}_{n\kappa }$ is a normalization constants. Also, the above
wave functions can be expressed in terms of the hypergeometric function as%
\begin{equation}
F_{n\kappa }(z)=\mathcal{N}_{n\kappa }z^{c_{12}}(1-c_{3}z)^{c_{13}}%
\begin{array}{c}
_{2}F_{1}%
\end{array}%
\left( -n,1+c_{10}+c_{11}+n;c_{10}+1;c_{3}z\right) ,
\end{equation}%
where $c_{12}>0,$ $c_{13}>0$ and $z\in \left[ 0,1/c_{3}\right] .$

$\label{appendix copy(1)}$

\section{Solving Quartic Energy Equation}

In order to solve the quartic equation (41b), the first step in the solution
is to define the following variables%
\begin{equation}
u=a_{2}-\frac{3}{8}a_{3}^{2};\text{ }v=a_{1}+\frac{1}{8}a_{3}^{3}-\frac{1}{2}%
a_{3}a_{2};\text{ }w=a_{0}-\frac{3}{256}a_{3}^{4}+\frac{1}{16}a_{3}^{2}a_{2}-%
\frac{1}{4}a_{3}a_{1};
\end{equation}%
which enable us to write down the related auxiliary cubic equation of the
form: 
\begin{equation}
a\overline{E}_{n\kappa }^{3}+b\overline{E}_{n\kappa }^{2}+c\overline{E}%
_{n\kappa }+d=0,\text{ }
\end{equation}%
with coefficients%
\begin{equation}
a=1,\text{ }b=\frac{u}{2},\text{ }c=\frac{1}{16}\left( u^{2}-4w\right) ,%
\text{ }d=-\frac{v^{2}}{64}.
\end{equation}%
The next step is solving the cubic equation (B2) by defining the variables%
\begin{equation}
f=\frac{c}{a}-\frac{1}{3}\frac{b^{2}}{a^{2}};\text{ }g=\frac{2}{27}\frac{%
b^{3}}{a^{3}}-\frac{1}{3}\frac{bc}{a^{2}}+\frac{d}{a};\text{ }h=\frac{1}{4}%
g^{2}+\frac{1}{27}f^{3},
\end{equation}%
and then applying one of the following three cases:

(i) When all three roots are real ($h<0$): we define the variables%
\begin{equation}
s=\sqrt{\frac{1}{4}g^{2}-h};\text{ }k=\cos ^{-1}\left( -\frac{g}{2s}\right) ,%
\text{ }F=\frac{1}{\sqrt[3]{s}},\text{ }H=\cos \left( \frac{k}{3}\right) ,%
\text{ }G=\sqrt{3}\sin \left( \frac{k}{3}\right) ,
\end{equation}%
where all the arguments in the trigonometric functions are in radians, to
obtain the three possible roots of (B2) in the form%
\begin{equation}
\overline{E}_{n\kappa }^{(1)}=2\frac{H}{F}-\frac{1}{3}\frac{b}{a};\text{ }%
\overline{E}_{n\kappa }^{(2)}=F\left( H+G\right) +\frac{3a}{b};\text{ }%
\overline{E}_{n\kappa }^{(3)}=F\left( H-G\right) +\frac{3a}{b},
\end{equation}%
and hence the four energies of the original quartic equation (41b) take the
forms%
\begin{equation*}
E_{n\kappa }^{(1)}=\sqrt{\overline{E}_{n\kappa }^{(2)}}+\sqrt{\overline{E}%
_{n\kappa }^{(3)}}-\frac{g}{8\sqrt{\overline{E}_{n\kappa }^{(2)}\overline{E}%
_{n\kappa }^{(3)}}}-\frac{1}{4}\frac{a_{3}}{a_{4}};\text{ }E_{n\kappa
}^{(2)}=\sqrt{\overline{E}_{n\kappa }^{(2)}}-\sqrt{\overline{E}_{n\kappa
}^{(3)}}+\frac{g}{8\sqrt{\overline{E}_{n\kappa }^{(2)}\overline{E}_{n\kappa
}^{(3)}}}-\frac{1}{4}\frac{a_{3}}{a_{4}};
\end{equation*}%
\begin{equation}
E_{n\kappa }^{(3)}=-\sqrt{\overline{E}_{n\kappa }^{(2)}}+\sqrt{\overline{E}%
_{n\kappa }^{(3)}}+\frac{g}{8\sqrt{\overline{E}_{n\kappa }^{(2)}\overline{E}%
_{n\kappa }^{(3)}}}-\frac{1}{4}\frac{a_{3}}{a_{4}};\text{ }E_{n\kappa
}^{(4)}=-\sqrt{\overline{E}_{n\kappa }^{(2)}}-\sqrt{\overline{E}_{n\kappa
}^{(3)}}-\frac{g}{8\sqrt{\overline{E}_{n\kappa }^{(2)}\overline{E}_{n\kappa
}^{(3)}}}-\frac{1}{4}\frac{a_{3}}{a_{4}}.
\end{equation}%
It is worth noting that whenever we have three real root we always choose
the two non-zero roots; say, $\overline{E}_{n\kappa }^{(2)}$ and $\overline{E%
}_{n\kappa }^{(3)}$ in (B6) of the cubic equation.

(ii) When only one root is real ($h>0$): the definitions 
\begin{equation}
R=-\frac{1}{2}g+\sqrt{h};\text{ }S=\sqrt[3]{R};\text{ }T=-\frac{1}{2}g-\sqrt{%
h};\text{ }U=\sqrt[3]{T};
\end{equation}%
enable us to write down the three roots of (B2):%
\begin{equation*}
\overline{E}_{n\kappa }^{(1)}=S+U-\frac{b}{3a};\text{ }\overline{E}_{n\kappa
}^{(2)}=-\frac{1}{2}\left( S+U\right) -\frac{b}{3a}+i\frac{\sqrt{3}}{2}%
\left( S-U\right) ;
\end{equation*}%
\begin{equation}
\overline{E}_{n\kappa }^{(3)}=-\frac{1}{2}\left( S+U\right) -\frac{b}{3a}-i%
\frac{\sqrt{3}}{2}\left( S-U\right) ,
\end{equation}%
and hence the four energies of Eq. (41b) are%
\begin{equation*}
E_{n\kappa }^{(1)}=\sqrt{\overline{E}_{n\kappa }^{(2)}}+\sqrt{\overline{E}%
_{n\kappa }^{(3)}}-\frac{g}{8\sqrt{\overline{E}_{n\kappa }^{(2)}\overline{E}%
_{n\kappa }^{(3)}}}-\frac{1}{4}\frac{a_{3}}{a_{4}};\text{ }E_{n\kappa
}^{(2)}=\sqrt{\overline{E}_{n\kappa }^{(2)}}-\sqrt{\overline{E}_{n\kappa
}^{(3)}}+\frac{g}{8\sqrt{\overline{E}_{n\kappa }^{(2)}\overline{E}_{n\kappa
}^{(3)}}}-\frac{1}{4}\frac{a_{3}}{a_{4}};
\end{equation*}%
\begin{equation}
E_{n\kappa }^{(3)}=-\sqrt{\overline{E}_{n\kappa }^{(2)}}+\sqrt{\overline{E}%
_{n\kappa }^{(3)}}+\frac{g}{8\sqrt{\overline{E}_{n\kappa }^{(2)}\overline{E}%
_{n\kappa }^{(3)}}}-\frac{1}{4}\frac{a_{3}}{a_{4}};\text{ }E_{n\kappa
}^{(4)}=-\sqrt{\overline{E}_{n\kappa }^{(2)}}-\sqrt{\overline{E}_{n\kappa
}^{(3)}}-\frac{g}{8\sqrt{\overline{E}_{n\kappa }^{(2)}\overline{E}_{n\kappa
}^{(3)}}}-\frac{1}{4}\frac{a_{3}}{a_{4}}.
\end{equation}%
It is worth noting that whenever we have one real root and two complex roots
we always choose the two complex roots.

(iii) When all three roots are real and equal ($f=g=h=0$), then the roots of
(B2):%
\begin{equation}
\overline{E}_{n\kappa }^{(1)}=\overline{E}_{n\kappa }^{(2)}=\overline{E}%
_{n\kappa }^{(3)}=-\sqrt[3]{\frac{d}{a}},
\end{equation}%
and hence the four energies of the original quartic equation (41b) are%
\begin{equation}
E_{n\kappa }^{(1)}=-2\sqrt[3]{\frac{d}{a}}-\frac{1}{4}\frac{a_{3}}{a_{4}};%
\text{ }E_{n\kappa }^{(2)}=E_{n\kappa }^{(3)}=-\frac{1}{4}\frac{a_{3}}{a_{4}}%
;\text{ }E_{n\kappa }^{(4)}=2\sqrt[3]{\frac{d}{a}}-\frac{1}{4}\frac{a_{3}}{%
a_{4}}.
\end{equation}

$\label{appendix}$

\section{Normalization constants}

The normalization constant, $\mathcal{N}_{nl}$ can be determined in closed
form. We start by using the relation between the hypergeometric function and
the Jacobi polynomials (see formula (8.962.1) in [49]): 
\begin{equation*}
\begin{array}{c}
_{2}F_{1}%
\end{array}%
\left( -n,n+\nu +\mu +1;\nu +1;\frac{1-x}{2}\right) =\frac{n!}{\left( \nu
+1\right) _{n}}P_{n}^{\left( \nu ,\mu \right) }(x),
\end{equation*}%
\begin{equation}
\left( \nu +1\right) _{n}=\frac{\Gamma (n+\nu +1)}{\Gamma (\nu +1)},
\end{equation}%
to rewrite the wave functions in (32) as%
\begin{equation}
F_{n\kappa }(r)=\mathcal{N}_{n\kappa }\frac{n!\Gamma (2\varepsilon _{n\kappa
}+1)}{\Gamma (n+2\varepsilon _{n\kappa }+1)}e^{-\varepsilon _{n\kappa
}\alpha r}(1-e^{-\alpha r})^{\delta _{1}}P_{n}^{(2\varepsilon _{n\kappa
},2\delta _{1}-1)}(1-2e^{-\alpha r}).
\end{equation}%
From the normalization condition $\int_{0}^{\infty }\left[ u_{n,l}(r)\right]
^{2}dr=1$ and under the coordinate change $x=1-2e^{-\alpha r},$ the
normalization constant in (B2) is given by%
\begin{equation}
\mathcal{N}_{n\kappa }^{-2}=\frac{1}{\alpha }\left[ \frac{n!\Gamma
(2\varepsilon _{n\kappa }+1)}{\Gamma (n+2\varepsilon _{n\kappa }+1)}\right]
^{2}\int_{-1}^{1}\left( \frac{1-x}{2}\right) ^{2\varepsilon _{n\kappa
}}\left( \frac{1+x}{2}\right) ^{2\delta _{1}-1}\left( \frac{1+x}{2}\right) %
\left[ P_{n}^{(2\varepsilon _{n\kappa },2\delta _{1}-1)}(x)\right] ^{2}dx.
\end{equation}%
The calculation of this integral can be done by writting 
\begin{equation*}
\frac{1+x}{2}=1-\left( \frac{1-x}{2}\right) ,
\end{equation*}%
and using the following two integrals (see formula (7.391.5) in [49]):%
\begin{equation}
\int_{-1}^{1}\left( 1-x\right) ^{\nu -1}\left( 1+x\right) ^{\mu }\left[
P_{n}^{\left( \nu ,\mu \right) }(x)\right] ^{2}dx=2^{\nu +\mu }\frac{\Gamma
(n+\nu +1)\Gamma (n+\mu +1)}{n!\nu \Gamma (n+\nu +\mu +1)},
\end{equation}%
which is valid for $\func{Re}$($\nu )>0$ and $\func{Re}$($\mu )>-1$ and (see
formula (7.391.1) in [49]):%
\begin{equation}
\int_{-1}^{1}\left( 1-x\right) ^{\nu }\left( 1+x\right) ^{\mu }\left[
P_{n}^{\left( \nu ,\mu \right) }(x)\right] ^{2}dx=2^{\nu +\mu +1}\frac{%
\Gamma (n+\nu +1)\Gamma (n+\mu +1)}{n!\Gamma (n+\nu +\mu +1)(2n+\nu +\mu +1)}%
,
\end{equation}%
which is valid for $\func{Re}$($\nu )>-1,$ $\func{Re}$($\mu )>-1.$ Thus, the
normalization constant:%
\begin{equation}
\mathcal{N}_{n\kappa }=\frac{1}{\Gamma (2\varepsilon _{n\kappa }+1)}\left[ 
\frac{\alpha \varepsilon _{n\kappa }(n+\varepsilon _{n\kappa }+\delta _{1})}{%
2(n+\delta _{1})}\frac{\Gamma (n+2\varepsilon _{n\kappa }+1)\Gamma
(n+2\varepsilon _{n\kappa }+2\delta _{1})}{n!\Gamma \left( n+2\delta
_{1}\right) }\right] ^{1/2},
\end{equation}%
where $0\leq n,\kappa <\infty .$ In the $s$-wave $\left( \kappa =-1\right) $
case, the above result is written explicitly as 
\begin{equation}
\mathcal{N}_{n,-1}=\frac{1}{\Gamma (2\eta +1)}\left[ \frac{\alpha \eta
(n+\eta +\delta )}{2(n+\delta )}\frac{\Gamma (n+2\eta +1)\Gamma (n+2\eta
+2\delta )}{n!\Gamma \left( n+2\delta \right) }\right] ^{1/2}.
\end{equation}%
Also, the non-relativistic normalization constant is therefore obtained as%
\begin{equation}
\mathcal{N}_{nl}=\frac{1}{\Gamma (2\widetilde{\eta }+1)}\left[ \frac{\alpha 
\widetilde{\eta }(n+\widetilde{\eta }+\widetilde{\delta })}{2(n+\widetilde{%
\delta })}\frac{\Gamma (n+2\widetilde{\eta }+1)\Gamma (n+2\widetilde{\eta }+2%
\widetilde{\delta })}{n!\Gamma \left( n+2\widetilde{\delta }\right) }\right]
^{1/2}.
\end{equation}

\newpage

{\normalsize 
}

\bigskip

\baselineskip= 2\baselineskip
\bigskip \newpage {\normalsize 
}

\baselineskip= 2\baselineskip
\bigskip \newpage \FRAME{ftbpFO}{0.0277in}{0.0277in}{0pt}{\Qct{The GMP model
with $D=15$ $fm^{-1}$ for (a) various potential ranges $\protect\alpha %
=0.05, $ $0.15,$ $0.30$ $fm^{-1}$ along with $r_{e}=0.40$ $fm,$ and (b)
various equilibrium inter-nuclear distances $r_{e}=0.40,$ $0.80$, $1.20$ $fm$
along with $\protect\alpha =0.40$ $fm^{-1}.$}}{}{Figure 1}{}\FRAME{ftbpFO}{%
0.0277in}{0.0277in}{0pt}{\Qct{The upper and lower spinor wave functions, in
the exact spin symmetry, for (a) $n=0$ $(0p_{1/2},0p_{3/2})$ spin doublet
eigenstates with $\protect\kappa =1$ and $\protect\kappa =-2$ and (b) $n=1$ $%
(1p_{1/2},1p_{3/2})$ spin doublet eigenstates with $\protect\kappa =1$ and $%
\protect\kappa =-2$.}}{}{Figure 2}{}

\bigskip

\begin{table}[tbp]
\caption{Specific values of the constants in the solution of the GMP under
spin symmetry.}%
\begin{tabular}{lll}
\tableline Constant &  & Constant \\ 
\tableline$c_{1}=1$ &  & $c_{2}=1$ \\ 
$c_{3}=1$ &  & c$_{4}=0$ \\ 
$c_{5}=-\frac{1}{2}$ &  & $c_{6}=\frac{1}{4}\left[ 1+4\left( \left(
2+b\right) \xi _{1}-\widetilde{E}_{n\kappa }\right) \right] $ \\ 
$c_{7}=$2$\left( \widetilde{E}_{n\kappa }-\xi _{1}\right) -\kappa \left(
\kappa +1\right) $ &  & $c_{8}=-\widetilde{E}_{n\kappa }$ \\ 
$c_{9}=\frac{1}{4}\left[ \left( 1+2\kappa \right) ^{2}+4b\xi _{1}\right] $ & 
& $c_{10}=2\eta _{1}=2i\sqrt{\widetilde{E}_{n\kappa }}$ \\ 
$c_{11}=\sqrt{\left( 1+2\kappa \right) ^{2}+4b\xi _{1}}$ &  & $c_{12}=\eta
_{1}$ \\ 
$c_{13}=\frac{1}{2}\left( 1+\sqrt{\left( 1+2\kappa \right) ^{2}+4b\xi _{1}}%
\right) $ &  & $A=\left( 2+b\right) \xi _{1}-\widetilde{E}_{n\kappa }$ \\ 
$B=-$2$\left( \widetilde{E}_{n\kappa }-\xi _{1}\right) -\kappa \left( \kappa
+1\right) $ &  & $C=-\widetilde{E}_{n\kappa }$ \\ 
\tableline &  & 
\end{tabular}%
\end{table}

\begin{table}[tbp]
\caption{Specific values of the constants in the solution of the GMP under
pseudospin symmetry.}%
\begin{tabular}{lll}
\tableline Constant &  & Constant \\ 
\tableline$c_{1}=1$ &  & $c_{2}=1$ \\ 
$c_{3}=1$ &  & c$_{4}=0$ \\ 
$c_{5}=-\frac{1}{2}$ &  & $c_{6}=\frac{1}{4}\left[ 1+4\left( \left(
2+b\right) \xi _{2}-\widetilde{E}_{n\kappa }\right) \right] $ \\ 
$c_{7}=$2$\left( \widetilde{E}_{n\kappa }-\xi _{2}\right) -\kappa \left(
\kappa -1\right) $ &  & $c_{8}=\eta _{2}^{2}=-\widetilde{E}_{n\kappa }$ \\ 
$c_{9}=\frac{1}{4}\left[ \left( 1-2\kappa \right) ^{2}+4b\xi _{2}\right] $ & 
& $c_{10}=2\eta _{2}=2i\sqrt{\widetilde{E}_{n\kappa }}$ \\ 
$c_{11}=\sqrt{\left( 1-2\kappa \right) ^{2}+4b\xi _{2}}$ &  & $c_{12}=\eta
_{2}$ \\ 
$c_{13}=\frac{1}{2}\left( 1+\sqrt{\left( 1-2\kappa \right) ^{2}+4b\xi _{2}}%
\right) $ &  & $A=\left( 2+b\right) \xi _{1}-\widetilde{E}_{n\kappa }$ \\ 
$B=-$2$\left( \widetilde{E}_{n\kappa }-\xi _{2}\right) -\kappa \left( \kappa
-1\right) $ &  & $C=\eta _{2}^{2}=-\widetilde{E}_{n\kappa }$ \\ 
\tableline &  & 
\end{tabular}%
\end{table}

\bigskip

\bigskip 
\begin{table}[tbp]
\caption{The Schr\"{o}dinger bound state energy levels $E_{nl}$ (in au) of
the GMP as functions of $\protect\alpha $ and $r_{e}$ for various states
with $D=15$ where $\hbar =\protect\mu =1.$}%
\begin{tabular}{ll}
\begin{tabular}{llllllll}
\tableline &  & $r_{e}=0.4$ &  &  & $r_{e}=0.8$ &  &  \\ 
States & $\alpha $ & Present & DG [53] & LS [54] & Present & DG [53] & LS
[54] \\ 
\tableline$2p$ & $0.05$ & $7.86080$ & $7.8606$ & $7.8628$ & $4.14088$ & $%
4.14068$ & $4.14208$ \\ 
& $0.10$ & $7.95329$ & $7.95247$ & $7.95537$ & $4.21917$ & $4.21835$ & $%
4.2204$ \\ 
& $0.15$ & $8.04508$ & $8.04322$ & $8.04724$ & $4.29737$ & $4.29552$ & $%
4.2987$ \\ 
& $0.20$ & $8.13616$ & $8.13287$ & $8.13842$ & $4.37551$ & $4.37221$ & $%
4.3769$ \\ 
& $0.25$ & $8.22656$ & $8.22142$ & $8.22892$ & $4.45360$ & $4.44845$ & $%
4.4551$ \\ 
& $0.30$ & $8.31630$ & $8.30889$ & $8.31874$ & $4.53166$ & $4.52425$ & $%
4.5332$ \\ 
$3p$ & $0.05$ & $10.9978$ & $10.9976$ & $10.9998$ & $7.53279$ & $7.53258$ & $%
7.5350$ \\ 
& $0.10$ & $11.1626$ & $11.1617$ & $11.1647$ & $7.72475$ & $7.72393$ & $%
7.7271$ \\ 
& $0.15$ & $11.3242$ & $11.3224$ & $11.32647$ & $7.91516$ & $7.9133$ & $%
7.9177$ \\ 
& $0.20$ & $11.4828$ & $11.4795$ & $11.48513$ & $8.10400$ & $8.10071$ & $%
8.1066$ \\ 
& $0.25$ & $11.6383$ & $11.6331$ & $11.64068$ & $8.29129$ & $8.28615$ & $%
8.2941$ \\ 
& $0.30$ & $11.7907$ & $11.7833$ & $11.67565$ & $8.47703$ & $8.46962$ & $%
8.4799$ \\ 
$3d$ & $0.05$ & $10.2160$ & $10.2154$ & $10.21651$ & $5.73974$ & $5.73913$ & 
$5.7404$ \\ 
& $0.10$ & $10.3535$ & $10.351$ & $10.35409$ & $5.84574$ & $5.84327$ & $%
5.8465$ \\ 
& $0.15$ & $10.4893$ & $10.4837$ & $10.48992$ & $5.95061$ & $5.94505$ & $%
5.9515$ \\ 
& $0.20$ & $10.6233$ & $10.6135$ & $10.62403$ & $6.05441$ & $6.04453$ & $%
6.0553$ \\ 
& $0.25$ & $10.7557$ & $10.7403$ & $10.75645$ & $6.15720$ & $6.14177$ & $%
6.1582$ \\ 
& $0.30$ & $10.8864$ & $10.8642$ & $10.88719$ & $6.25904$ & $6.23682$ & $%
6.2601$ \\ 
$4p$ & $0.05$ & $12.4976$ & $12.4974$ & $12.4992$ & $9.61301$ & $9.6128$ & $%
9.6156$ \\ 
& $0.10$ & $12.6968$ & $12.696$ & $12.69851$ & $9.88351$ & $9.88269$ & $%
9.8862$ \\ 
& $0.15$ & $12.8883$ & $12.8865$ & $12.8901$ & $10.1485$ & $10.1467$ & $%
10.1514$ \\ 
& $0.20$ & $13.0722$ & $13.0689$ & $13.0740$ & $10.4080$ & $10.4047$ & $%
10.4111$ \\ 
& $0.25$ & $13.2484$ & $13.2433$ & $13.2501$ & $10.6619$ & $10.6568$ & $%
10.665$ \\ 
$4d$ & $0.05$ & $12.0983$ & $12.0977$ & $12.0989$ & $8.49334$ & $8.49272$ & $%
8.4948$ \\ 
& $0.10$ & $12.2850$ & $12.2825$ & $12.2857$ & $8.70708$ & $8.70461$ & $%
8.7087$ \\ 
& $0.15$ & $12.4664$ & $12.4608$ & $12.46715$ & $8.91774$ & $8.91218$ & $%
8.9194$ \\ 
& $0.20$ & $12.6424$ & $12.6326$ & $12.64324$ & 9.12538 & $9.11551$ & $%
9.1272 $%
\end{tabular}
& 
\begin{tabular}{llllllll}
\tableline &  & $r_{e}=0.4$ &  &  & $r_{e}=0.8$ &  &  \\ 
States & $\alpha $ & Present & DG [53] & LS [54] & Present & DG [53] & LS
[54] \\ 
\tableline$4f$ & $0.05$ & $11.8208$ & $11.8195$ & $11.8209$ & $7.43469$ & $%
7.43346$ & $7.4351$ \\ 
& $0.10$ & $11.9979$ & $11.993$ & $11.9981$ & $7.58636$ & $7.58142$ & $%
7.5868 $ \\ 
& $0.15$ & $12.1716$ & $12.1604$ & $12.1718$ & $7.73559$ & $7.72448$ & $%
7.7361$ \\ 
& $0.20$ & $12.3418$ & $12.3221$ & $12.3421$ & $7.88251$ & $7.86276$ & $%
7.8831$ \\ 
$5p$ & $0.10$ & $13.5421$ & $13.5413$ & $13.5434$ & $11.3021$ & $11.3012$ & $%
11.3047$ \\ 
& $0.20$ & $13.9289$ & $13.9257$ & $13.9301$ & $11.9132$ & $11.9099$ & $%
11.9161$ \\ 
$5d$ & $0.10$ & $13.3068$ & $13.3043$ & $13.3075$ & $10.5201$ & $10.5176$ & $%
10.5219$ \\ 
& $0.20$ & $13.6925$ & $13.6827$ & $13.6931$ & $11.0692$ & $11.0594$ & $%
11.0713$ \\ 
$5f$ & $0.10$ & $13.1475$ & $13.1426$ & $13.1478$ & $9.7966$ & $9.79166$ & $%
9.7975$ \\ 
& $0.20$ & $13.5332$ & $13.5134$ & $13.5333$ & $10.2728$ & $10.253$ & $%
10.2738$ \\ 
$5g$ & $0.10$ & $13.0379$ & $13.0296$ & $13.0379$ & $9.15212$ & $9.14389$ & $%
9.1524$ \\ 
& $0.20$ & $13.4267$ & $13.3938$ & $13.42667$ & $9.55246$ & $9.51954$ & $%
9.5528$ \\ 
$6p$ & $0.10$ & $14.0521$ & $14.0513$ & $14.0530$ & $12.2798$ & $12.279$ & $%
12.2822$ \\ 
$6d$ & $0.10$ & $13.9070$ & $13.9045$ & $13.9075$ & $11.7364$ & $11.7339$ & $%
11.7383$ \\ 
$6f$ & $0.10$ & $13.8111$ & $13.8062$ & $13.8113$ & $11.2448$ & $11.2398$ & $%
11.2459$ \\ 
$6g$ & $0.10$ & $13.7465$ & $13.7383$ & $13.7466$ & $10.8152$ & $10.807$ & $%
10.8158$ \\ 
\tableline &  &  &  &  &  &  & 
\end{tabular}%
\end{tabular}%
\end{table}
\begin{table}[tbp]
\caption{The Schr\"{o}dinger bound state energy levels $E_{nl}$ (in au) of
the Kratzer potential as a function of $r_{e}$ for various states with $D=15$
where $\hbar =\protect\mu =1.$}%
\begin{tabular}{llllll}
\tableline State/$r_{e}=$ & $0.1$ & $0.4$ & $0.8$ & $1.0$ & $1.5$ \\ 
\tableline$2p$ & $13.9765$ & $7.76759$ & $4.06249$ & $3.21339$ & $2.07749$
\\ 
$3p$ & $14.5308$ & $10.8298$ & $7.33925$ & $6.26836$ & $4.56776$ \\ 
$3d$ & $14.5192$ & $10.0766$ & $5.63256$ & $4.41694$ & $2.74032$ \\ 
$4p$ & $14.7319$ & $12.2908$ & $9.33707$ & $8.27299$ & $6.40188$ \\ 
$4d$ & $14.7269$ & $11.9062$ & $8.27643$ & $7.04418$ & $5.05045$ \\ 
$4f$ & $14.7246$ & $11.6401$ & $7.28048$ & $5.81633$ & $3.61468$ \\ 
$5p$ & $14.8268$ & $13.0996$ & $10.6443$ & $9.65890$ & $7.79162$ \\ 
$5d$ & $14.8242$ & $12.8774$ & $9.94062$ & $8.80199$ & $6.76420$ \\ 
$5f$ & $14.8230$ & $12.7278$ & $9.29975$ & $7.96875$ & $5.69315$ \\ 
$5g$ & $14.8224$ & $12.6242$ & $8.73653$ & $7.18597$ & $4.60125$ \\ 
$6p$ & $14.8789$ & $13.5937$ & $11.5460$ & $10.6568$ & $8.86976$ \\ 
$6d$ & $14.8774$ & $13.4539$ & $11.0555$ & $10.0356$ & $8.07050$ \\ 
$6f$ & $14.8768$ & $13.3616$ & $10.6191$ & $9.44444$ & $7.25037$ \\ 
$6g$ & $14.8764$ & $13.2985$ & $10.2430$ & $8.89964$ & $6.42694$ \\ 
\tableline &  &  &  &  & 
\end{tabular}%
\end{table}

\bigskip

\bigskip 
\begin{table}[tbp]
\caption{The eigeenergies of the Dirac valence states in units of $fm^{-1}$
for several values of $n$ and $\protect\kappa $ with the parameters $M=1.0$ $%
fm^{-1}$ and $D=15$ $fm^{-1}$ in the case of exact spin symmetry limit ($%
C_{s}=0$ $fm^{-1}$). We have set $\hbar =c=1.$}%
\begin{tabular}{lllllll}
\tableline$l$ & $n,\kappa <0$ & $nL_{j=l+1/2}$ & $E_{n,\kappa <0}$ & $%
n,\kappa >0$ & $nL_{j=l-1/2}$ & $E_{n,\kappa >0}$ \\ 
\tableline$\alpha =0.10$ $fm^{-1},$ & $r_{e}=0.40$ $fm$ &  &  &  &  &  \\ 
$1$ & $0,-2$ & $0p_{3/2}$ & $5.5791076$ & $0,1$ & $0p_{1/2}$ & $5.5791076$
\\ 
$2$ & $0,-3$ & $0d_{5/2}$ & $6.8118605$ & $0,2$ & $0d_{3/2}$ & $6.8118605$
\\ 
$3$ & $0,-4$ & $0f_{7/2}$ & $8.0171073$ & $0,3$ & $0f_{5/2}$ & $8.0171073$
\\ 
$4$ & $0,-5$ & $0g_{9/2}$ & $9.1025175$ & $0,4$ & $0g_{7/2}$ & $9.1025175$
\\ 
$1$ & $1,-2$ & $1p_{3/2}$ & $8.1823677$ & $1,1$ & $1p_{1/2}$ & $8.1823677$
\\ 
$2$ & $1,-3$ & $1d_{5/2}$ & $8.8815340$ & $1,2$ & $1d_{3/2}$ & $8.8815340$
\\ 
$3$ & $1,-4$ & $1f_{7/2}$ & $9.6603105$ & $1,3$ & $1f_{5/2}$ & $9.6603105$
\\ 
$4$ & $1,-5$ & $1g_{9/2}$ & $10.4200196$ & $1,4$ & $1g_{7/2}$ & $10.4200196$
\\ 
\tableline$\alpha =0.30$ $fm^{-1},$ & $r_{e}=0.40$ $fm$ &  &  &  &  &  \\ 
$1$ & $0,$ $-2$ & $0p_{3/2}$ & $5.7078594$ & $0,1$ & $0p_{1/2}$ & $5.7078594$
\\ 
$2$ & $0,-3$ & $0d_{5/2}$ & $6.9646771$ & $0,2$ & $0d_{3/2}$ & $6.9646771$
\\ 
$3$ & $0,-4$ & $0f_{7/2}$ & $8.2121326$ & $0,3$ & $0f_{5/2}$ & $8.2121326$
\\ 
$4$ & $0,-5$ & $0g_{9/2}$ & $9.3506414$ & $0,4$ & $0g_{7/2}$ & $9.3506414$
\\ 
$1$ & $1,-2$ & $1p_{3/2}$ & $8.4626850$ & $1,1$ & $1p_{1/2}$ & $8.4626850$
\\ 
$2$ & $1,-3$ & $1d_{5/2}$ & $9.1762544$ & $1,2$ & $1d_{3/2}$ & $9.1762544$
\\ 
$3$ & $1,-4$ & $1f_{7/2}$ & $9.9831712$ & $1,3$ & $1f_{5/2}$ & $9.9831712$
\\ 
$4$ & $1,-5$ & $1g_{9/2}$ & $10.7812870$ & $1,4$ & $1g_{7/2}$ & $10.7812870$
\\ 
\tableline$\alpha =0.10$ $fm^{-1},$ & $r_{e}=0.80$ $fm$ &  &  &  &  &  \\ 
$1$ & $0,$ $-2$ & $0p_{3/2}$ & $3.6831690$ & $0,1$ & $0p_{1/2}$ & $3.6831690$
\\ 
$2$ & $0,-3$ & $0d_{5/2}$ & $4.3378367$ & $0,2$ & $0d_{3/2}$ & $4.3378367$
\\ 
$3$ & $0,-4$ & $0f_{7/2}$ & $5.0775317$ & $0,3$ & $0f_{5/2}$ & $5.0775317$
\\ 
$4$ & $0,-5$ & $0g_{9/2}$ & $5.8291637$ & $0,4$ & $0g_{7/2}$ & $5.8291637$
\\ 
$1$ & $1,-2$ & $1p_{3/2}$ & $5.8388616$ & $1,1$ & $1p_{1/2}$ & $5.8388616$
\\ 
$2$ & $1,-3$ & $1d_{5/2}$ & $6.2180443$ & $1,2$ & $1d_{3/2}$ & $6.2180443$
\\ 
$3$ & $1,-4$ & $1f_{7/2}$ & $6.6999906$ & $1,3$ & $1f_{5/2}$ & $6.6999906$
\\ 
$4$ & $1,-5$ & $1g_{9/2}$ & $7.2334098$ & $1,4$ & $1g_{7/2}$ & $7.2334098$
\\ 
\tableline &  &  &  &  &  & 
\end{tabular}%
\end{table}
\begin{table}[tbp]
\caption{The eigenenergies of the Dirac hole states in units of $fm^{-1}$
for several values of $n$ and $\protect\kappa $ with the parameters $M=1.0$ $%
fm^{-1}$ and $D=15$ $fm^{-1}$ in the case of the pseudospin limit. We have
set $\hbar =c=1.$}%
\begin{tabular}{lllllllll}
\tableline$\widetilde{l}$ & $n,\kappa <0$ & $\left( l,j\right) $ & $%
E_{n,\kappa <0}$ & $E_{n,\kappa <0}$ & $n-1,\kappa >0$ & $\left(
l+2,j+1\right) $ & $E_{n-1,\kappa >0}$ & $E_{n-1,\kappa >0}$ \\ 
\tableline$\alpha =0.10$ $fm^{-1},$ & $r_{e}=0.40$ $fm$ &  & $C_{ps}=0$ & $%
C_{ps}=5.0$ &  &  & $C_{ps}=0$ & $C_{ps}=5.0$ \\ 
$1$ & $1,-1$ & $1s_{1/2}$ & $7.1975980$ & $9.0681299$ & $0,2$ & $0d_{3/2}$ & 
$7.1975980$ & $9.0681299$ \\ 
$2$ & $1,-2$ & $1p_{3/2}$ & $7.9184956$ & $9.7118773$ & $0,3$ & $0f_{5/2}$ & 
$7.9184956$ & $9.7118773$ \\ 
$3$ & $1,-3$ & $1d_{5/2}$ & $8.6773995$ & $10.3429382$ & $0,4$ & $0g_{7/2}$
& $8.6773995$ & $10.3429382$ \\ 
$4$ & $1,-4$ & $1f_{7/2}$ & $9.3887565$ & $10.9088876$ & $0,5$ & $0h_{9/2}$
& $9.3887565$ & $10.9088876$ \\ 
$1$ & $2,-1$ & $2s_{1/2}$ & $8.7054710$ & $10.1707692$ & $1,2$ & $1d_{3/2}$
& $8.7054710$ & $10.1707692$ \\ 
$2$ & $2,-2$ & $2p_{3/2}$ & $9.1768973$ & $10.6114449$ & $1,3$ & $1f_{5/2}$
& $9.1768973$ & $10.6114449$ \\ 
$3$ & $2,-3$ & $2d_{5/2}$ & $9.7101038$ & $11.0728353$ & $1,4$ & $1g_{7/2}$
& $9.7101038$ & $11.0728353$ \\ 
$4$ & $2,-4$ & $2f_{7/2}$ & $10.2352574$ & $11.5024429$ & $1,5$ & $1h_{9/2}$
& $10.2352574$ & $11.5024429$ \\ 
\tableline$\alpha =0.30$ $fm^{-1},$ & $r_{e}=0.40$ $fm$ &  & $C_{ps}=0$ & $%
C_{ps}=-5.0$ &  &  & $C_{ps}=0$ & $C_{ps}=-5.0$ \\ 
$1$ & $1,-1$ & $1s_{1/2}$ & $7.4924218$ & $6.2513272$ & $0,2$ & $0d_{3/2}$ & 
$7.4924218$ & $6.2513272$ \\ 
$2$ & $1,-2$ & $1p_{3/2}$ & $8.2327097$ & $6.9536060$ & $0,3$ & $0f_{5/2}$ & 
$8.2327097$ & $6.9536060$ \\ 
$3$ & $1,-3$ & $1d_{5/2}$ & $9.0263198$ & $7.7574142$ & $0,4$ & $0g_{7/2}$ & 
$9.0263198$ & $7.7574142$ \\ 
$4$ & $1,-4$ & $1f_{7/2}$ & $9.7816289$ & $8.5599024$ & $0,5$ & $0h_{9/2}$ & 
$9.7816289$ & $8.5599024$ \\ 
$1$ & $2,-1$ & $2s_{1/2}$ & $9.1118917$ & $8.0615654$ & $1,2$ & $1d_{3/2}$ & 
$9.1118917$ & $8.0615654$ \\ 
$2$ & $2,-2$ & $2p_{3/2}$ & $9.5965089$ & $8.5254559$ & $1,3$ & $1f_{5/2}$ & 
$9.5965089$ & $8.5254559$ \\ 
$3$ & $2,-3$ & $2d_{5/2}$ & $10.1539358$ & $9.0879925$ & $1,4$ & $1g_{7/2}$
& $10.1539358$ & $9.0879925$ \\ 
$4$ & $2,-4$ & $2f_{7/2}$ & $10.7110039$ & $9.6762585$ & $1,5$ & $1h_{9/2}$
& $10.7110039$ & $9.6762585$ \\ 
\tableline$\alpha =0.10$ $fm^{-1},$ & $r_{e}=0.80$ $fm$ &  & $C_{ps}=0$ & $%
C_{ps}=-10.0$ &  &  & $C_{ps}=0$ & $C_{ps}=-10.0$ \\ 
$1$ & $1,-1$ & $1s_{1/2}$ & $4.9972498$ & $2.8369925$ & $0,2$ & $0d_{3/2}$ & 
$4.9972498$ & $2.8369925$ \\ 
$2$ & $1,-2$ & $1p_{3/2}$ & $5.4281366$ & $3.1294486$ & $0,3$ & $0f_{5/2}$ & 
$5.4281366$ & $3.1294486$ \\ 
$3$ & $1,-3$ & $1d_{5/2}$ & $5.9435202$ & $3.5245459$ & $0,4$ & $0g_{7/2}$ & 
$5.9435202$ & $3.5245459$ \\ 
$4$ & $1,-4$ & $1f_{7/2}$ & $6.4876355$ & $3.9879463$ & $0,5$ & $0h_{9/2}$ & 
$6.4876355$ & $3.9879463$ \\ 
$1$ & $2,-1$ & $2s_{1/2}$ & $6.3778620$ & $4.3337279$ & $1,2$ & $1d_{3/2}$ & 
$6.3778620$ & $4.3337279$ \\ 
$2$ & $2,-2$ & $2p_{3/2}$ & $6.6705985$ & $4.5479115$ & $1,3$ & $1f_{5/2}$ & 
$6.6705985$ & $4.5479115$ \\ 
$3$ & $2,-3$ & $2d_{5/2}$ & $7.0449439$ & $4.8443318$ & $1,4$ & $1g_{7/2}$ & 
$7.0449439$ & $4.8443318$ \\ 
$4$ & $2,-4$ & $2f_{7/2}$ & $7.4613765$ & $5.2011464$ & $1,5$ & $1h_{9/2}$ & 
$7.4613765$ & $5.2011464$ \\ 
\tableline &  &  &  &  &  &  &  & 
\end{tabular}%
\end{table}

\end{document}